\documentclass[10pt,a4paper]{article}

\usepackage[english]{babel}
\usepackage[utf8]{inputenc} 
\usepackage[inner=3cm,outer=3cm,bottom=3cm,top=3cm]{geometry}
\usepackage{graphicx}
\usepackage{array}
\usepackage{morefloats}
\usepackage{amsfonts,amsmath,color}
\usepackage{xcolor}												
\usepackage{tikz}
\usepackage[T1]{fontenc}    
\usepackage{appendix}
\usepackage{stackrel}
\usepackage{eurosym}
\usepackage{subfigure}
\usepackage{epstopdf} 
\usepackage{amscd,amssymb,amsthm,latexsym, amsbsy,amsmath}
\usepackage{multicol}
\usepackage{multirow}
\usepackage{booktabs}
\usepackage{float}
\usepackage{color, colortbl}
\usepackage{xcolor}
\usepackage{enumerate} 
\usepackage[round]{natbib} 
\usepackage{rotating}  
\usepackage{textcomp} 
\usepackage{longtable} 
\usepackage{authblk}
\usepackage{mathtools} 
\usepackage{arydshln} 
\usepackage[hidelinks]{hyperref} 

\newtheorem{theorem}{Theorem}[section]
\newtheorem{algorithm}[theorem]{Algorithm}				

\newcommand{\Var}{\mathbb{V}}
\newcommand{\Cov}{\mathbb{C}}
\newcolumntype{T}[1]{>{\centering\let\newline\\\arraybackslash\hspace{0pt}}m{#1}}
\LTcapwidth=\textwidth

\providecommand{\keywords}[1]{\textbf{\textit{Key words---}} #1}

\title{Novel specification tests for additive concurrent model formulation based on martingale difference divergence}
\vspace{0.1cm}
\author[1,*]{Laura Freijeiro-González}
\author[1]{Manuel Febrero-Bande}
\author[1]{Wenceslao González-Manteiga}
\affil[1]{Centre for Mathematical Research and Technology Transfer of Galicia (CITMAga). Department of Statistics, Mathematical Analysis and Optimization. Universidade de Santiago de Compostela, Santiago de Compostela, Spain. *Email: laura.freijeiro.gonzalez@usc.es }
\date{}                     
\begin{document}
	

\maketitle

\begin{abstract}	
	Novel significance tests are proposed for the quite general additive concurrent model formulation without the need of model, error structure preliminary estimation or the use of tuning parameters. Making use of the martingale difference divergence coefficient, we propose new tests to measure the conditional mean independence in the concurrent model framework taking under consideration all observed time instants. In particular, global dependence tests to quantify the effect of a group of covariates in the response as well as partial ones to apply covariates selection are introduced. Their asymptotic distribution is obtained on each case and a bootstrap algorithm is proposed to compute its p-values in practice. These new procedures are tested by means of simulation studies and some real datasets analysis.
\end{abstract}	



\section{Introduction}

A general concurrent model is a regression model where the response
$Y\in \mathbb{R}$ and $p\geq1$ covariates $X=(X_1,\dots, X_p)\in \mathbb{R}^p$ are all functions of the same argument $t\in \mathcal{D}_t$, and the influence is concurrent, simultaneous or point-wise in the sense that $X$ is assumed to only influence $Y(t)$ through its value $X(t)=\left(X_1(t),\dots,X_p(t) \right)\in \mathbb{R}^p$ at time $t$ by means of the relation
\begin{equation}\label{multi_conc_mod}
	Y(t)=m(t,X(t))+\varepsilon(t),
\end{equation}
where $m(\cdot)$ is an unknown function collecting the $\mathbb{E}\left[ Y(t)|_{X(t)} \right]$ information and $\varepsilon(t)$ is the error of the model. This last is a process which is assumed to have mean zero, independent of $X$ and with covariance function $\Omega(s,t)=\text{cov}\left(\varepsilon(s),\varepsilon(t)\right)$.\\

The concurrent model displayed in (\ref{multi_conc_mod}) is in the middle of longitudinal and functional data. This would depend on the number of observed time instants in the $t$ domain $\mathcal{D}_t$. When this number is dense enough, we can treat our sample data as curves, which translates in a functional data framework. Otherwise, if time instants are not dense and these are quite spaced respective to the $t$ real domain, the longitudinal framework would be more adequate. It is still an open problem to determine the inflection point between both situations. We refer the reader to the work of \cite{wang2017unified} for a discussion about this topic.\\

There are a lot of contexts where the (\ref{multi_conc_mod}) formulation arises both in functional or longitudinal framework form. In any situation where data can be monitored, as in health, environmental or financial issues among others, this model can be employed. Some examples can be seen in works as the ones of \cite{Xue2007} or \cite{JiangWang2011} for the longitudinal data context, exemplifying by means of epidemiology studies of AIDS data sets. Other real data examples in medicine can be found in \cite{goldsmith2017variable} or \cite{wang2017unified}. There, making use of the concurrent model formulation, a blood pressure study to detect masked hypertension is carried out in the first case and a United States flu data study as well as Alzheimer’s disease neuroimaging record data are modeled in the second one. More examples about health and nutrition applying concurrent models are displayed in \cite{Kim2018Additive} or \cite{GHOSAL2022Score} by means of studies related with gait deficiency, dietary calcium absorption and relation between child mortality and financial power in different countries. Examples in the environmental field are collected in the works of \cite{JunZhang2011} or \cite{Ospina2019} based on describing forest nitrogen cycling and modeling the rainfall ground respectively. A completely different example is the work of \cite{Ghosal2022}, where casual bike rentals in Washington, D.C., are explained concurrently by means of meteorological variables. All these examples bring out the fact that the concurrent model is a very transversal and wide-employed tool nowadays.\\

Nevertheless, an inconvenient of the concurrent model general formulation displayed in (\ref{multi_conc_mod}) is that the $m(\cdot)$ structure is quite difficult to be estimated in practice without any assumption about its form. In literature, it is quite common to assume linearity, which translates in assuming that $m(t,X(t))=\beta(t)X(t)$ in (\ref{multi_conc_mod}), and work under this premise. However, this assumption can be quite restrictive in practice and more general structures are needed to properly model real examples. This last results in a gain in flexibility but adds complexity to the estimation procedure. A discussion about the effort done in estimating concurrent model different structures is done in \cite{maity2017nonparametric}. This highlights the fact that more information is needed to proper estimate $m(\cdot)$. In conclusion, it is important to certainly guarantee that there exists useful information on the covariates $X$ to model the mean of $Y$ as a preliminary step. Besides, covariates selection algorithms for the concurrent model are of interest to avoid irrelevant covariates and simplify the estimation process.\\

As a result, a first step to assure the veracity of the model structure displayed in (\ref{multi_conc_mod}) is to verify if all $p$ covariates $\{X_1(t),\dots, X_p(t)\}$ contribute to the correct explanation of $Y(t)$ or some can be excluded from the model formulation. For this purpose, taking $D\subset \{1,\dots,p\}$, a dependence test can be performed by means of testing
\begin{equation*}
	\begin{split}
		&H_0: \mathbb{E} \left[ Y(t) |_{X_D(t)} \right]=\mathbb{E} \left[ Y(t) \right] \quad \text{almost surely} \; \forall t\in \mathcal{D}_t \setminus \mathcal{N}_t\\
		&H_a: \mathbb{P}\left(  \mathbb{E} \left[ Y(t) |_{X_D(t)} \right] \neq \mathbb{E} \left[ Y(t) \right] \right)>0 \quad \forall t\in \mathcal{V}_t
	\end{split}
\end{equation*}
where $X_D(t)$ denotes the subset of $X(t)$ considering only the covariates with index in $D$, $\mathcal{D}_t \setminus \mathcal{N}_t$ is the domain of $t$ minus a null set $\mathcal{N}_t \subset \mathcal{D}_t$ and $\mathcal{V}_t \subset \mathcal{D}_t$ is a positive measure set.\\

However, quoting \cite{Zhang2018}, the above problem is very challenging in practice without assuming any structure of $m(\cdot)$. This is due to the huge class of alternatives targeted, related with growing dimension and nonlinear dependence. To solve this drawback, they propose to test the nullity of main effects first, keeping a type of hierarchical order. In this way, we test if additive and separate effects first enter the model before considering interaction terms. This results in the new test displayed in (\ref{exp_contrast}).

\begin{equation}\label{exp_contrast}
	\begin{split}
		&H_0: \mathbb{E} \left[ Y(t) |_{X_j(t)} \right]=\mathbb{E} \left[ Y(t) \right] \quad \text{almost surely} \; \forall  t\in \mathcal{D}_t \setminus \mathcal{N}_t \; \text{and every}\; j\in D\\
		&H_a: \mathbb{P}\left(  \mathbb{E} \left[ Y(t) |_{X_j(t)} \right] \not= \mathbb{E} \left[ Y(t) \right] \right)>0 \quad \forall t\in \mathcal{V}_t \; \text{and some}\; j\in D
	\end{split}
\end{equation}

Then, the rejection of the null hypothesis of (\ref{exp_contrast}) automatically implies the rejection of $H_0: \mathbb{E} \left[ Y(t) |_{X_D(t)} \right]=\mathbb{E}\left[ Y(t) \right]$ hypothesis. It is important to highlight that the reciprocal is not always true. In this way, the model (\ref{multi_conc_mod}) only makes sense if we can reject the $H_0$ hypothesis of (\ref{exp_contrast}). Otherwise, the covariates does not supply relevant information to explain $Y$. It is notorious that formulation (\ref{exp_contrast}) collects a wide range of dependence structures between $X$ and $Y$ in terms of additive regression models, with $m\left(t,X(t) \right)=F_1\left(t,X_1(t) \right)+\dots+F_p\left(t,X_p(t) \right)$. Moreover, it is not necessary to known the real form of $m(\cdot)$ to determine if the effect of $X$ is significant or not.\\

To the best of our knowledge, there is not literature about significance tests for the additive concurrent model avoiding previous estimation of its effects or extra parameters. We refer to \cite{wang2017unified} and \cite{GHOSAL2022Score} for these in the linear formulation. They propose effect tests over the $\beta(t)$ function making use of the empirical likelihood. Thus, they provide tools to test if all $p$ covariates are relevant or some of them can be excluded from the model once its parameters are estimated in the linear framework. Nevertheless, a proper effects estimation involving several tuning parameters values to estimate as well as the linearity hypothesis are necessary to guarantee its adequate performance. In terms of the $\beta(t)$ structure estimation, \cite{wang2017unified} propose to make use of a local linear estimator, which depends on a proper bandwidth selection, whereas \cite{GHOSAL2022Score} employs a finite basis function expansion, which requires the selection of the number of considered basis terms. Besides, this last procedure needs to estimate the error model structure, which translates in an extra functional basis representation and extra parameters estimation. All of this translates in difficulties in the estimation procedure even if the linear hypothesis can be accepted. Currently, \cite{Kim2018Additive} developed a new significance test in a more general framework to alleviate the linear hypothesis assumption: additive effects are considered  in the concurrent model formulation displayed in (\ref{multi_conc_mod}) instead of linear ones. In this work, F-test techniques are employed over a functional basis representation of the additive effects to detect relevant covariates. Again, this technique depends on a good preliminary estimation of the model effects to be able to select relevant covariates by means of significance tests. The correct selection of the number of basis functions for each considered covariate/effect representation is still an open problem in this framework, where these quantities play the role of tuning parameters. Furthermore, a proper error variance estimation is needed to standardize the covariates as an initial step. As this structure is unknown in practice, \cite{Kim2018Additive} assume this can be decomposed as a sum of a zero-mean smooth stochastic process plus a zero-mean white noise measurement error with variance $\sigma^2$, resulting in the autocovariance function $\Omega(s,t)=\Sigma(s,t)+\sigma^2 \mathbb{I}\{s=t\}$, and work under this premise. Nevertheless, this assumption can be restrictive in practice. In consequence, significance tests without any assumption in the model structure as well as the necessity of a preliminary estimation step are desirable.\\

Other procedures with a different approach, but with the common objective of selecting covariates, are the implementation of Bayesian selectors or penalization techniques in the concurrent model estimation process. We can highlight the works of \cite{goldsmith2017variable} or \cite{Ghosal2020} in the linear formulation and the one of \cite{Ghosal2022} for general additive effects. While \cite{goldsmith2017variable} use the spike-and-slab regression covariates selection procedure, \cite{Ghosal2020} and \cite{goldsmith2017variable} implement penalizations based on LASSO (\cite{Tibshirani1996}), SCAD (\cite{fan2001variable}), MCP (\cite{Zhang2010}) or its grouped versions (\cite{yuan2006model}), respectively. As a result, covariates selection is implemented at same time as estimation. However, some tuning parameters are needed in all these methodologies: we need to determine the number of basis functions to represent the effects in all of them jointly with prior parameters in case of the spike-and-slab regression or the amount of penalization otherwise. As a result, prior estimation of parameters is needed in this approach too.\\

In this paper, we deal with this concern bridging a gap for significance tests without previous model estimation and being able to assess the usefulness of $X$ for modeling the expectation of $Y$ in a pretty general formulation. Besides, our approach avoids extra tuning parameters estimation as well as the need of modeling the error structure. For this aim, we propose a novel statistic for the concurrent model based on the martingale difference divergence ideas of \cite{martingale2014} to perform (\ref{exp_contrast}). As a result, this tests if there exists effect of the covariates in the conditional mean explanation of $Y$ no matter the underlying form of $m(\cdot)$ assuming additive effects.\\

It is important to notice that we can consider $D=\{1,\dots,p\}$ to perform (\ref{exp_contrast}), which translates in testing if all $p$ covariates are relevant, or only a subset $D\subset \{1,\dots,p\}$ with cardinality $1\leq d<p$. In this last case, we test if only a bunch of covariates are relevant, excluding the rest from the model. A special case is to consider $D=\{j\}$ for some $j=1,\dots,p$. This approach allows us to implement covariates screening with no need to estimate the regressor function. In fact, we can test the effect of every covariate in this way. This results in $j=1,\dots,p$ partial tests of the form
\begin{equation}\label{exp_j_contrast}
	\begin{split}
		&H_{0j}: \mathbb{E} \left[ Y(t) |_{X_j(t)} \right]=\mathbb{E} \left[ Y(t) \right] \quad \text{almost surely} \; \forall t\in \mathcal{D}_t \setminus \mathcal{N}_t\\
		&H_{aj}: \mathbb{P}\left(  \mathbb{E} \left[ Y(t) |_{X_j(t)} \right] \neq \mathbb{E} \left[ Y(t) \right] \right)>0 \quad \forall t\in \mathcal{V}_t
	\end{split}
\end{equation}

Thus, we can test if a small subset of $\{1,\dots,p\}$ is suitable to fit the additive concurrent model or whether all covariates need to be considered. As a result, we can avoid noisy covariates entering the model and reduce the problem dimension.\\

The rest of the paper is organized as follows. In Section \ref{sec:MDD} the martingale difference divergence coefficient is introduced along with some remarkable properties. We propose our new dependence tests in Section \ref{sec:test_MDD}, theoretically justifying their good behavior and proposing a bootstrap scheme to calculate its p-values in practice. A simulation study to test their performance is presented in Section \ref{sec:simu_scenarios}, jointly, a comparison with the \cite{GHOSAL2022Score} and \cite{Kim2018Additive} techniques performance is displayed. Next, we apply our proposed tests to three real datasets in Section \ref{sec:real_data}. Eventually, some discussion arise in Section \ref{sec:conclusions}.

\section{Martingale difference divergence (MDD)}\label{sec:MDD}

The martingale difference divergence (MDD) was introduced by \cite{martingale2014}. This coefficient is a natural extension of the covariance distance of \cite{Szekely2007}, \cite{szekely2017energy} but to measure the departure from conditional mean independence between a scalar response variable $Y\in\mathbb{R}$ and a vector predictor $X\in\mathbb{R}^p$. Hence, this idea can be used to screen out continuous variables that do not contribute to the conditional mean of a regression model response given the covariates. This translates in the test 

\begin{equation}\label{martingale_cont}
	\begin{split}
		& H_0:\mathbb{E}[Y|_X]=\mathbb{E}[Y]\;\text{almost surely}	\\
		& H_a: \mathbb{P} \left( \mathbb{E}[Y|_X]\not=\mathbb{E}[Y] \right) >0
	\end{split}
\end{equation}

Therefore, a coefficient measuring the difference between the conditional mean and the unconditional one is needed to perform (\ref{martingale_cont}). Following similar ideas and argumentation of the correlation distance measure of \cite{Szekely2007}, it emerges the MDD.\\

Then, for $X\in\mathbb{R}^p$ and $Y\in\mathbb{R}$, the MDD of $Y$ given $X$ is the nonnegative number $MDD(Y|_X)$ defined by
	\begin{equation}\label{MDD}
		MDD^2(Y|_X)=\frac{1}{c_p} \int_{\mathbb{R}^p} \frac{|g_{Y,X}(s)-g_Yg_X(s)|}{|s|_p^{p+1}} ds
	\end{equation}
where $c_p=\pi^{(1+p)/2}/ \Gamma((1+p)/2)$, $g_{Y,X}(s)=\mathbb{E}[Y e^{i<s,X>}]$, $g_Y=\mathbb{E}[Y]$ and $g_X(s)=\mathbb{E}[ e^{i<s,X>}]$, being $i=\sqrt{-1}$ the imaginary unit, $<\cdot,\cdot>$ the inner product, $|\cdot|$ the norm of the complex function space, defined by $|f|=f \bar{f}$ for $f(\cdot)$ a complex value function with conjugate $\bar{f}$ and $|\cdot|_p$ is the euclidean norm of the $\mathbb{R}^p$ space.\\

It can be seen in \cite{martingale2014} or \cite{ParkShaoYao2015} that, being $(X',Y')$ and $(X'',Y'')$ independent copies of $(X,Y)$, an alternative way to the definition (\ref{MDD}) is 
\begin{equation}\label{MDD_UV}
	\begin{split}
		MDD^2(Y|_X)=& \mathbb{E}\left[ K(X,X')L(Y,Y') \right]+\mathbb{E}\left[ K(X,X')\right] \mathbb{E}\left[L(Y,Y') \right] \\
		& -2\mathbb{E}\left[ K(X,X')L(Y,Y'') \right]\\
		=& -\mathbb{E}\left[ \left( Y-\mathbb{E}[Y] \right) \left( Y'-\mathbb{E}[Y'] \right) |X-X'|_p \right]
	\end{split}	
\end{equation}
where $L(y,y')=(y-y')^2/2$ and $K(x,x')=|x-x'|_p$.\\

Since in general $MDD(Y|_X)\not=MDD(X|_Y)$, this is named divergence instead of distance. The MDD equals $0$ if and only if it is verified the $H_0$ hypothesis of (\ref{martingale_cont}) and otherwise MDD>0. Therefore, we can rewrite the test (\ref{martingale_cont}) as the new one displayed in (\ref{martingale_cont_rew}).

\begin{equation}\label{martingale_cont_rew}
	\begin{split}
		& H_0:MDD^2(Y|_X)=0 \;\text{almost surely}	\\
		& H_a: P\left( MDD^2(Y|_X)\not=0\right)>0
	\end{split}
\end{equation}

Next, we present an unbiased estimator of MDD introduced in \cite{Zhang2018}. Then, given $n$ observations $(X_i,Y_i)_{i=1}^n$ from the joint distribution of $(X,Y)$ with $X_i=(X_{i1},\dots,X_{ip})^\top\in \mathbb{R}^p$ and $Y_i\in \mathbb{R}$ we can define $A=(A_{il})_{i,l=1}^n$ and $B=(B_{il})_{i,l=1}^n$, where $A_{il}=|X_i-X_l|_p$ and $B_{il}=|Y_i-Y_l|^2/2$ for $i,l=1,\dots,n$. Following the $\mathcal{U}$-centered ideas of \cite{ParkShaoYao2015} we can define the $\mathcal{U}$-centered versions of $A$ and $B$, $\overline{A}$ and $\overline{B}$ respectively, given by
\begin{equation*}
	\begin{split}
		\overline{A}_{il}&= A_{il}-\frac{1}{n-2}\sum_{q=1}^{n} A_{iq} -\frac{1}{n-2} \sum_{q=1}^{n} A_{ql} + \frac{1}{(n-1)(n-2)}\sum_{q,r=1}^{n} A_{qr}\\
		\overline{B}_{il}&= B_{il}-\frac{1}{n-2}\sum_{q=1}^{n} B_{iq} -\frac{1}{n-2} \sum_{q=1}^{n} B_{ql} + \frac{1}{(n-1)(n-2)}\sum_{q,r=1}^{n} B_{qr}
	\end{split}
\end{equation*}

As a result, an unbiased estimator for MDD is defined by
	\begin{equation}\label{MDD_unbias}
		MDD_n^2(Y|_X)=( \overline{A} \cdot \overline{B}) = \frac{1}{n(n-3)}\sum_{i\neq l} \overline{A}_{il}\overline{B}_{il}.
	\end{equation}
A proof that $MDD_n^2(Y|_X)$ is an unbiased estimator for $MDD^2(Y|_X)$ can be found in Section 1.1 of the supplementary material of \cite{Zhang2018}.\\

An important characteristic of the $MDD_n^2(Y|X)$ unbiased estimator defined in (\ref{MDD_unbias}) is that this is a $\mathcal{U}$-statistic of order four. In fact, with some calculation, it can be proved that 
\begin{equation} \label{MDD_U_stat}
	MDD_n^2(Y|_X)=\frac{1}{\binom{n}{4}} \sum_{i<l<q<r} h(Z_i,Z_l,Z_q,Z_r)
\end{equation}
with 
\begin{equation*}
	\begin{split}
	h(Z_i,Z_l,Z_q,Z_r)&=\frac{1}{4!} \sum_{(s,w,u,v)}^{(i,l,q,r)}(A_{sw}B_{uv}+A_{sw}B_{sw}-2A_{sw}B_{su})\\
	&=\frac{1}{6} \sum_{s<w,u<v}^{(i,l,q,r)} (A_{sw}B_{uv}+A_{sw}B_{sw}) - \frac{1}{12} \sum_{(s,w,u)}^{(i,l,q,r)} A_{sw}B_{su}		
	\end{split}
\end{equation*}
where $Z_i=(X_i,Y_i)$ for $i=1,\dots,n$ and the summation is over all permutation of the 4-tuples of indices $(i,l,q,r)$. A guidelines about this calculation is provided in Section 1.1 of the supplementary material of \cite{Zhang2018}.\\

In view of the (\ref{MDD_U_stat}) formulation with $h(\cdot)$ a symmetric function, we can directly notice that $MDD_n^2(Y|_X)$ is a $\mathcal{U}$-statistic of order four by proper definition. Then, this statistic can be employed to perform test (\ref{martingale_cont}).

\section{Significance tests based on MDD}\label{sec:test_MDD}

Once we have a tool to measure conditional mean independence between $Y\in\mathbb{R}$ and a vector $X=(X_1,\dots,X_p)^\top\in\mathbb{R}^p$, we adapt this idea to the concurrent model case. For this aim, we make use of the ideas presented in \cite{Zhang2018} work for the vectorial framework. Henceforth, we assume a situation where all curves points are observed at same time instants. We start considering how to apply this procedure for a fixed time instant $t_u\in\mathcal{D}_t$, which corresponds with the vectorial framework. Then, we extend this methodology for a global dependence test considering all the domain $\mathcal{D}_t$.\\

Then, given an instant $t_u\in\mathcal{D}_t$, we observe $n_u \geq 1$ samples of the form $\{Y(t_u),X(t_u)\}_{u=1}^{n_u}$ with $Y(t_u)\in\mathbb{R}$ and $X(t_u)=(x_1(t_u),\dots,x_p(t_u))^\top\in\mathbb{R}^p$. Thus, we are under the vectorial framework assumptions of \cite{Zhang2018}. Rewriting the dependence test introduced in (\ref{martingale_cont}) to this framework we obtain the test
\begin{equation*}
	\begin{split}
	&H_0: \mathbb{E}[Y(t_u)|_{X_j(t_u)}]=\mathbb{E}[Y(t_u)] \;\text{almost surely for all }j=1,\dots,p \\
	&Ha: P\left( \mathbb{E}[Y(t_u)|_{X_j(t_u)}]\neq\mathbb{E}[Y(t_u)] \right)>0 \;\text{for some }j=1,\dots,p
	\end{split}
\end{equation*}
which can be expressed in terms of the MDD as
\begin{equation*}
	\begin{split}
	& H_0:MDD^2(Y(t_u)|_{X_j(t_u)})=0 \;\text{almost surely for all }j=1,\dots,p \\
	& H_a: P\left( MDD^2(Y(t_u)|_{X_j(t_u)})\not=0\right)>0\;\text{for some }j=1,\dots,p
	\end{split}
\end{equation*}

Making use of the unbiased estimator of the MDD introduced in (\ref{MDD_unbias}), \cite{Zhang2018} propose the statistic
\begin{equation}\label{T_n_vec}
	T_{n_u}=\sqrt{\binom{n_u}{2}}\frac{\sum_{j=1}^{p}MDD_{n_u}^2(Y(t_u)|_{X_j(t_u)})}{\hat{\mathcal{S}}_u}
\end{equation}
where $\hat{\mathcal{S}}_u^2$ is a suitable variance estimator of the theoretical variance $\mathcal{S}_u^2$ at instant $t_u\in\mathcal{D}_t$. This is given by\\

\begin{equation*}
	\hat{\mathcal{S}}_u^2= \frac{2}{n_u(n_u-1)c_{n_u}} \sum_{1\leq l<q\leq n_u} \sum_{j,j'=1}^{p} \left( \overline{A}_{lq} (t_u)\right)_j \left( \overline{A}_{lq} (t_u)\right)_{j'} \overline{B}^2_{lq}(t_u)
\end{equation*}

where

\begin{equation}\label{c_n}
	c_{n_u}= \frac{(n_u-3)^4}{(n_u-1)^4}+\frac{2(n_u-3)^4}{(n_u-1)^4(n_u-2)^3}+\frac{2(n_u-3)}{(n_u-1)^4(n_u-2)^3} \approx \frac{(n_u-3)^4}{(n_u-1)^4}.
\end{equation}

We refer to \cite{Zhang2018} for more details.\\

In order to study the asymptotic properties of $T_{n_u}$ we define $Z'(t_u)=\big( X'(t_u), Y'(t_u) \big)$ and $Z''(t_u)=\big( X''(t_u), Y''(t_u) \big)$ independent copies of $Z(t_u)=\big(X(t_u), Y(t_u)\big)$. Let $\dot{U}\big( X(t_u), X'(t_u) \big)=\sum_{j=1}^{p} U_j \big( x_j(t_u), x'_j (t_u) \big)$, being $U_j \big( x(t_u), x' (t_u) \big)= \mathbb{E} \big[ K\big(x_j(t_u),X'_j(t_u) \big) \big] + \mathbb{E} \big[ K\big(X_j(t_u),x'_j(t_u) \big) \big] - K\big(x_j(t_u),x'_j(t_u) \big) - \mathbb{E} \big[ K\big(X_j(t_u),X'_j(t_u) \big) \big] $, and $H\big(Z(t_u), Z'(t_u) \big)=V(Y(t_u), Y'(t_u))\dot{U}(X(t_u), X'(t_u))$, for $V(Y(t_u), Y'(t_u))= (Y(t_u)- \mathbb{E}[Y(t_u)])(Y'(t_u)- \mathbb{E}[Y(t_u)])$. Further define $G(Z(t_u),Z'(t_u))=\mathbb{E}\left[ H(Z(t_u),Z''(t_u))H(Z'(t_u),Z''(t_u))|_{(Z(t_u),Z'(t_u))} \right]$. Next, applying the theory developed by \cite{hall1980martingale} to our context, we proof that under $H_0$ and some assumptions, we can verified that $T_{n_u} \longrightarrow^d N(0,1)$ when $\hat{\mathcal{S}}_u^2/\mathcal{S}_u^2\longrightarrow^p 1$.

\begin{theorem}\label{th_conv_Tn}
	Under the assumption of $H_0$ and verifying
	\begin{equation*}
	\begin{split}
		\frac{\mathbb{E} \big[ G(Z(t_u),Z'(t_u))^2 \big]}{\big\{ \mathbb{E} \big[ H(Z(t_u),Z'(t_u))^2 \big] \big\}^2} &\longrightarrow 0\\
		\frac{ \mathbb{E} \big[ H\big(Z(t_u),Z'(t_u)\big)^4 \big]/n_u +\mathbb{E} \big[ H\big(Z(t_u),Z''(t_u)\big)^2 H\big(Z'(t_u),Z''(t_u)\big)^2 \big]  }{n_u \big\{\mathbb{E} \big[ H\big(Z(t_u),Z'(t_u)\big)^2 \big] \big\}^2} & \longrightarrow 0 \\
		\frac{ \mathbb{E} \big[ \dot{U}\big(X(t_u),X''(t_u)\big)^2 V\big(Y(t_u),Y'(t_u)\big)^2 \big] }{\mathcal{S}^2_u}&= o(n_u)\\
		\frac{\Var\big[Y(t_u)\big]^2 \sum_{j,j'=1}^{p} \text{dcov}\big(X_j(t_u), X_{j'}(t_u) \big)^2}{\mathcal{S}^2_u}&= o(n_u^2)\\
	\end{split}
	\end{equation*}
	for $\Var[\cdot]$ the variance operator and $\text{dcov}(\cdot,\cdot)$ the distance covariance,
	it is guarantee that $T_{n_u} \longrightarrow^d N(0,1)$ when $n_u\longrightarrow \infty$ and $\hat{\mathcal{S}}_u^2/\mathcal{S}_u^2\longrightarrow^p 1$.
\end{theorem}

We refer the reader to \cite{Zhang2018} for a more detailed explanation and deeper analysis of the required conditions.\\

Although we have guaranteed the asymptotic normality of our statistic, the convergence tends to be quite slow in practice. In view of this inconvenient, a wild bootstrap scheme is proposed, specially for the small sample size case. This is displayed in Algorithm \ref{wild_boot}.\\

\pagebreak
\begin{algorithm}[Wild bootstrap scheme using MDD]\label{wild_boot}\
	\begin{enumerate}
		\item{Generate the sample $\{e_{i_u}\}_{i_u=1}^{n_u}$ where $e_{i_u}$ are i.i.d. N(0,1).}
		\item{For every $j=1,\dots,p$ define the bootstrap $MDD^{*2}_{n_u}(Y(t_u)|_{X_j(t_u)})$ version as
			\begin{equation*}
				MDD^{*2}_{n_u}(Y(t_u)|_{X_j(t_u)})=\frac{1}{n_u(n_u-1)} \sum_{l\neq q} \left( \overline{A}_{lq} (t_u)\right)_j \overline{B}_{lq}(t_u) e_l e_q
			\end{equation*}
			where $\left( \overline{A}_{lq} (t_u)\right)_j$ and $\overline{B}_{lq}(t_u)$ are the $\mathcal{U}$-centered versions of $\left( A_{lq}(t_u) \right)_j=|X_{lj}(t_u)-X_{qj}(t_u)|$ and $B_{lq}(t_u)=|Y_l(t_u)-Y_q(t_u)|^2/2$, respectively.}
		\item{Calculate the bootstrap variance estimator 
			\begin{equation*}
				\hat{\mathcal{S}}^{*2}_u=\frac{1}{\binom{n_u}{2}} \sum_{1\leq l<q\leq n_u} \sum_{j,j'=1}^{p} \left( \overline{A}_{lq} (t_u)\right)_j \left( \overline{A}_{lq} (t_u)\right)_{j'} \overline{B}^2_{lq}(t_u) e^2_l e^2_q
		\end{equation*}}
		\item{Obtain the bootstrap statistic given by
			\begin{equation*}
				T_{n_u}^*=\sqrt{\binom{n_u}{2}}\frac{\sum_{j=1}^{p}MDD_{n_u}^{*2}(Y(t_u)|_{X_j(t_u)})}{\hat{\mathcal{S}}^*_u}
		\end{equation*}}
		\item{Repeat steps 1-4 a number $B$ of times obtaining $\{T_{n_u}^{*(1)},\dots, T_{n_u}^{*(B)}\}$. Calculate the bootstrap p-value as $\frac{1}{B} \sum_{b=1}^{B} \mathbb{I}\{T_{n_u}^{*(b)} \geq T_{n_u}\}$ being $\mathbb{I}(\cdot)$ the indicator function.}
	\end{enumerate}
\end{algorithm}

The good performance of the wild bootstrap scheme showed in Algorithm \ref{wild_boot} is proved in \cite{Zhang2018} under both, null and local alternatives. They also provide more insight about the $\hat{\mathcal{S}}^{*2}_u$ estimation.\\

Next, we extend these ideas to global and partial dependence tests in the concurrent model framework taking under consideration not one but all observed time instants.\\

\subsection{MDD global significance test} \label{subsec:global_test_MDD}

\begin{figure}[htb]
	\centering
	\includegraphics[width=\textwidth]{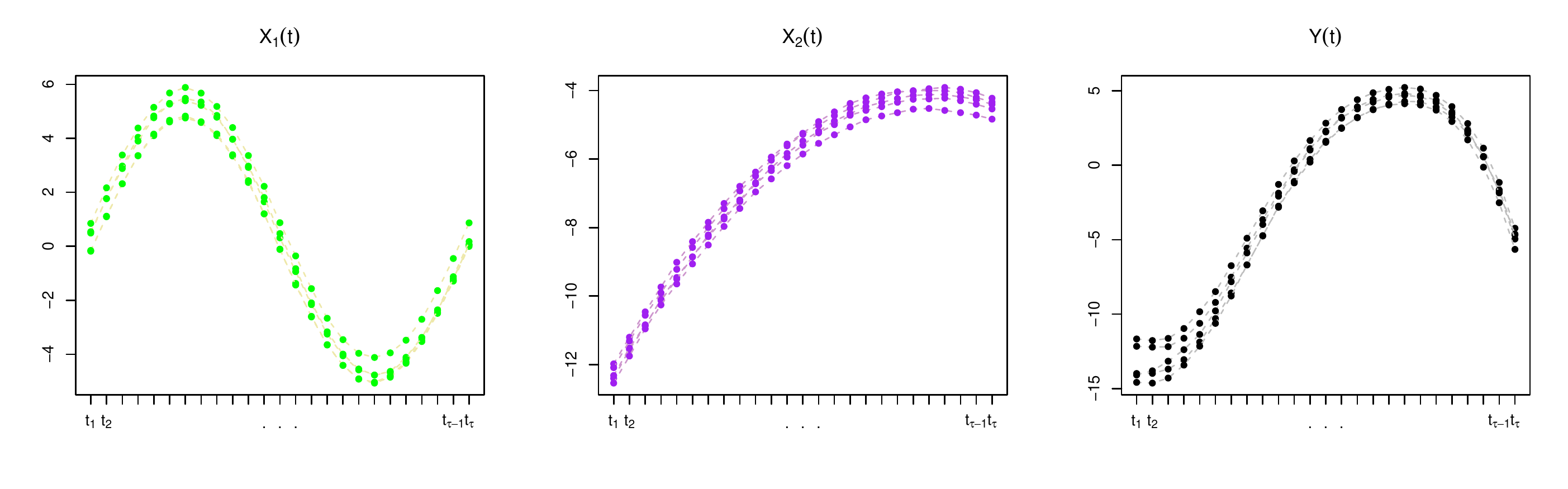}
	\caption{Example of a sample of five curves measured at same time instants $\{t_u\}_{u=1}^{\mathcal{T}}\in\mathcal{D}_t$ considering $p=2$ covariates ($X_1(t)$ and $X_2(t)$) to explain $Y(t)$. Filled points simulate a total of $n_u=3$ observed points at each instant $t_u$.}
	\label{example_param_same}
\end{figure}

Now, a total of $\{t_u\}_{u=1}^{\mathcal{T}}\in\mathcal{D}_t$ time instants are considered and there are $n_u$ observed samples, each of them of the form $\{Y_{i_u}(t_u),X_{i_u}(t_u)\}_{i_u=1}^{n_u}$. As we mentioned before, all curves are assumed to be observed at same time instants, which translates in $n_u=n$ for all $u=1,\dots,\mathcal{T}$. A graphic example of our current situation considering $p=2$ covariates for a concurrent model with structure similar to (\ref{multi_conc_mod}) is displayed in Figure \ref{example_param_same}.\\ 

In this way, we want to include all the information provided by the observed time instants $\{t_u\}_{u=1}^{\mathcal{T}}\in\mathcal{D}_t$ in a new statistic. Besides, as we mentioned before, we can be interested in testing dependence not only considering all covariates but a subset $D\subset \{1,\dots,p\}$. As a result, a global dependence test is applied over the complete trajectory considering the information provided by $D$. Rewriting (\ref{exp_contrast}), this gives place to the test
\begin{equation}\label{global_martingale_cont_rew}
	\begin{split}
		& H_0: \int_{\mathcal{D}_t\setminus \mathcal{N}_t} MDD^2(Y(t)|_{X_j(t)}) dt=0 \;\text{almost surely} \; \text{for every}\; j\in D	\\
		& H_a: \mathbb{P}\left( \int_{ \mathcal{V}_t} MDD^2(Y(t)|_{X_j(t)}) dt\not=0\right)>0 \; \text{for some}\; j\in D
	\end{split}
\end{equation}

In order to implement the new test introduced in (\ref{global_martingale_cont_rew}) a proper estimator of $\int_{\mathcal{D}_t} MDD^2(Y(t)|_{X_j(t)}) dt$ for every $j\in D$ is needed. For this purpose, we propose an integrated statistic based on
\begin{equation}\label{T}
	T_D= \sqrt{\binom{n}{2}}\frac{ \sum_{j\in D}  \int_{\mathcal{D}_t} MDD_{n}^2(Y(t)|_{X_j(t)})dt}{\widehat{\widetilde{\mathcal{S}}}_D} = \sqrt{\binom{n}{2}}\frac{ \sum_{j\in D} \widetilde{MDD}_{n}^2(Y(t)|_{X_j(t)})}{\widehat{\widetilde{\mathcal{S}}}_D},
\end{equation}
being $\widetilde{MDD}_{n}^2(Y(t)|_{X_j(t)})= \int_{\mathcal{D}_t} MDD_{n}^2(Y(t)|_{X_j(t)})dt$ and 
\begin{equation}\label{S_int}
	\widehat{\widetilde{\mathcal{S}}}_D^2=\frac{2}{n(n-1)c_n} \sum_{1\leq l<q\leq n} \sum_{j,j'\in D} \int_{\mathcal{D}_t} \left( \overline{A}_{lq} (t)\right)_j \left( \overline{A}_{lq} (t)\right)_{j'} \overline{B}^2_{lq}(t) dt
\end{equation}
a suitable variance estimator of $\sum_{j\in D} \widetilde{MDD}_{n}^2(Y(t)|_{X_j(t)})$ with $c_n$ defined as in (\ref{c_n}). See Section \ref{subsubsec:S_int} for in-depth details.\\

The $\widetilde{MDD}_{n}^2(Y(t)|_{X_j(t)})$ remains a $\mathcal{U}$-statistic of order four. This is because, denoting $Z_{ij}(t)=\left( X_{ij}(t), Y_i(t) \right)$ and $ (  \widetilde{A_{sw}B_{uv}} )_j=\int_{\mathcal{D}_t} \left( A_{sw}(t) \right)_j \left( B_{uv}(t)\right)_j dt$ for all $(s,w,u,v)$ we have that
\begin{equation}\label{h_tilde}
	\begin{split}
	\widetilde{h(Z_{ij}(t),Z_{lj}(t),Z_{qj}(t),Z_{rj}(t))}&=\int_{\mathcal{D}_t} h(Z_{ij}(t),Z_{lj}(t),Z_{qj}(t),Z_{rj}(t)) dt\\
	&=\frac{1}{4!} \sum_{(s,w,u,v)}^{(i,l,q,r)} \left\lbrace  \left( \widetilde{A_{sw}B_{uv}} \right)_j +  \left( \widetilde{A_{sw}B_{sw}} \right)_j -2 \left( \widetilde{A_{sw}B_{su}} \right)_j \right\rbrace \\
	&=\frac{1}{6} \sum_{s<w,u<v}^{(i,l,q,r)} \left\lbrace  \left( \widetilde{A_{sw}B_{uv}} \right)_j +  \left( \widetilde{A_{sw}B_{sw}} \right)_j \right\rbrace - \frac{1}{12} \sum_{(s,w,u)}^{(i,l,q,r)} \left( \widetilde{A_{sw}B_{su}} \right)_j		
\end{split}
\end{equation}
keeps being a measurable and symmetric function. Then, it is easy to see that we can write
\begin{equation*}
	\widetilde{MDD}_{n}^2(Y(t)|_{X_j(t)})=\frac{1}{\binom{n}{4}} \sum_{i<l<q<r} \widetilde{h(Z_{ij}(t),Z_{lj}(t),Z_{qj}(t),Z_{rj}(t))} 
\end{equation*}
which keeps the structure of a U-statistic of order 4.\\

It can be proved that $\widetilde{MDD}_n^2(Y(t)|_{X_j(t)})$ is an unbiased estimator of $\widetilde{MDD}^2(Y(t)|_{X_j(t)})$ following similar steps as the ones of \cite{Zhang2018}. See Section \ref{append:unbiasedness} of the Appendix.\\

Furthermore, due to the $\mathcal{U}$-statistics properties, similar argumentation to the one of Section \ref{sec:test_MDD} can be applied to this context with the aim of guaranteeing that the $T_D$ statistic of (\ref{T}) is asymptotically normal. Now, we have to consider the integrated version which gives place to Theorem \ref{th_conv_Tn_int}. A proof of this theorem is given in Section \ref{append:normality} of the Appendix. For this, we have followed similar steps as the ones of Section 1.3 of the supplementary material of \cite{Zhang2018} but adapted to our context.\\

\begin{theorem}\label{th_conv_Tn_int}
	Under the assumption of $H_0$ and verifying
	\begin{equation*}
		\begin{split}
			\frac{\mathbb{E} \left[ \widetilde{G(Z(t),Z'(t))}^2 \right]}{\left\lbrace \mathbb{E} \left[ \widetilde{H(Z(t),Z'(t))}^2 \right] \right\rbrace^2} &\longrightarrow 0\\
			\frac{ \mathbb{E} \left[ \widetilde{H\big(Z(t),Z'(t)\big)}^4 \right]/n +\mathbb{E} \left[ \widetilde{H\big(Z(t),Z''(t)\big)}^2 \widetilde{H\big(Z'(t),Z''(t)\big)}^2 \right]  }{n \left\lbrace \mathbb{E} \left[ \widetilde{H(Z(t),Z'(t))}^2 \right] \right\rbrace^2} & \longrightarrow 0 \\
			\frac{ \mathbb{E} \left[ \widetilde{\dot{U}\big(X(t),X''(t)\big)}^2 \widetilde{V\big(Y(t),Y'(t)\big)}^2 \right] }{\widetilde{\mathcal{S}}_D^2}&= o(n)\\
			\frac{ \sum_{j,j'\in D} \int_{\mathcal{D}_t} \Var \big[Y(t)\big]^2 \text{dcov}\big(X_j(t), X_{j'}(t) \big)^2 dt}{\widetilde{\mathcal{S}}_D^2}&= o(n^2)\\
		\end{split}
	\end{equation*}
	for $\Var[\cdot]$ the variance operator and $\text{dcov}(\cdot,\cdot)$ the distance covariance, it is guarantee that $T_{D} \longrightarrow^d N(0,1)$ when $n\longrightarrow \infty$ and $\widehat{\widetilde{\mathcal{S}}}_D^2/\widetilde{\mathcal{S}}_D^2\longrightarrow^p 1$.
\end{theorem}

Nevertheless, the  asymptotic convergence of the $T_D$ statistic can be very slow in practice too. To solve this issue we approximate the p-value by means of the adaptation of the wild bootstrap scheme introduced above in Section \ref{sec:test_MDD}. The bootstrap scheme for the global dependence test is collected in Algorithm \ref{wild_boot_global}. The consistency of the proposed wild bootstrap scheme and variance estimator for the concurrent model case is omitted due to the document extension. However, the proof results from plugging the integrated version in the one of \cite{Zhang2018}, introduced in Section 1.6 of their supplementary material.\\

\begin{algorithm}[Wild bootstrap scheme for global dependence test using MDD]\label{wild_boot_global}\
	\begin{enumerate}
		\item{For $u=1\dots,\mathcal{T}$:
			\begin{enumerate}[\theenumi.\arabic{enumii}.]
				
				\item{Calculate
					\begin{equation*}
						(T_{u})_D=\sqrt{\binom{n}{2}}\sum_{j\in D} MDD_{n}^2(Y(t_u)|_{X_j(t_u)}).
				\end{equation*} }
				
				\item{Obtain
					\begin{equation*}
						(\hat{\mathcal{S}}_{u})_D= \sqrt{\frac{2}{n(n-1)c_n}\sum_{1\leq l<q\leq n} \sum_{j,j'\in D}\left( \overline{A}_{lq} (t_u)\right)_j \left(\overline{A}_{lq} (t_u)\right)_{j'} \overline{B}^2_{lq}(t_u)},
					\end{equation*}
					where $\left( \overline{A}_{lq} (t_u)\right)_j$ and $\overline{B}_{lq}(t_u)$ are the $\mathcal{U}$-centered versions of $\left( A_{lq}(t_u) \right)_j=|X_{lj}(t_u)-X_{qj}(t_u)|$ and $B_{lq}(t_u)=|Y_l(t_u)-Y_q(t_u)|^2/2$, respectively. }
				
				\item{Generate the sample $\{e_{i}\}_{i=1}^{n}$ where $e_{i}$ are i.i.d. N(0,1).}
				
				\item{Define the bootstrap $MDD^{*2}_{n}(Y(t_u)|_{X_j(t_u)})$ version as
					\begin{equation*}
						MDD^{*2}_{n}(Y(t_u)|_{X_j(t_u)})=\frac{1}{n(n-1)} \sum_{l\neq q} \left( \overline{A}_{lq} (t_u)\right)_j \overline{B}_{lq}(t_u) e_l e_q
				\end{equation*}}
			
				\item{Obtain the bootstrap statistic numerator
					\begin{equation*}
						(T_{u}^*)_D=\sqrt{\binom{n}{2}} \sum_{j\in D} MDD_{n}^{*2}(Y(t_u)|_{X_j(t_u)}).
				\end{equation*}}
				
				\item{Calculate the bootstrap variance estimator 
					\begin{equation*}
						(\hat{\mathcal{S}}^{*}_{u})_D= \sqrt{\frac{1}{\binom{n}{2}} \sum_{1\leq l<q\leq n} \sum_{j, j' \in D} \left( \overline{A}_{lq} (t_u)\right)_j \left( \overline{A}_{lq} (t_u)\right)_{j'} \overline{B}^2_{lq}(t_u) e^2_l e^2_q}.
				\end{equation*}}

				\item{Repeat steps 1.3-1.6 a number $B$ of times obtaining the sets $\{(T^*_{u})_D^{(1)},\dots, (T^*_{u})_D^{(B)}\}$
					and $\{(\hat{\mathcal{S}}^*_u)_D^{(1)},\dots, (\hat{\mathcal{S}}^*_u)_D^{(B)}\}$.}
				
		\end{enumerate}}
		
		\item{Approximate the sample statistic $\tilde{(E)}_D=\int_{\mathcal{D}_t} (T_{t})_D/(\hat{\mathcal{S}}_{t})_D dt$ value by means of numerical techniques using $\{(T_{1})_D,\dots,(T_{\mathcal{T}})_D\}$ and $\{(\hat{\mathcal{S}}_{1})_D,\dots,(\hat{\mathcal{S}}_{\mathcal{T}})_D\}$.}
		
		\item{For every $b=1,\dots,B$, approximate the bootstrap statistic $(\tilde{E}^{*})^{(b)}_D=\int_{\mathcal{D}_t} (T^*_{t})^{(b)}_D/(\hat{\mathcal{S}}^*_t)^{(b)}_D dt$ value by means of numerical techniques using $\{(T^{*}_{1})_D^{(b)},\dots,(T^{*}_{\mathcal{T}})_D^{(b)} \}$ and $\{(\hat{\mathcal{S}}_1^{*})_D^{(b)},\dots,(\hat{\mathcal{S}}_\mathcal{T}^{*})_D^{(b)} \}$.}  
				
		\item{Obtain the bootstrap p-value as $\frac{1}{B} \sum_{b=1}^{B} \mathbb{I}\{(\tilde{E}^{*})^{(b)}_D \geq (\tilde{E})_D\}$ being $\mathbb{I}(\cdot)$ the indicator function.}
	\end{enumerate}
\end{algorithm}

In terms of $D$, a special case is the situation where all covariates are considered, $D=\{1,\dots,p\}$. First of all, we must check if, at least, some covariates supply relevant information to model $Y$. Considering $D$ the set of all covariates indices, we can check this premise performing (\ref{global_martingale_cont_rew}). In case of having no evidences to reject the null hypothesis of conditional mean independence, it does not make sense to model $Y$ with the available information. Otherwise, if the conditional mean independence is discarded in this initial step, we can be interested in searching for an efficient subset of covariates to reduce the problem dimension.\\

Then, for a subset $D\subset \{1,\dots,p\}$ with cardinality $d$, $1\leq d<p$, we can test if these $d$ covariates play a role in terms of the concurrent regression model by means of (\ref{global_martingale_cont_rew}). If not, we can discard them and to reduce the problem dimensionality to $p-d$. In case we are interested in covariates screening one by one, which corresponds with the case where $D=\{j\}$, we can apply the $j=1,\dots,p$ tests displayed in (\ref{exp_j_contrast}). This results in $p$ consecutive partial tests for $j=1,\dots,p$ considering $H_{0j}: \mathbb{E} \left[ Y(t) |_{X_j(t)} \right]=\mathbb{E} \left[ Y(t) \right]$ almost surely $\forall t\in \mathcal{D}_t\setminus \mathcal{N}_t$ or equivalently $H_{0j}: \widetilde{MDD}^2 \left( Y(t) |_{X_j(t)} \right)=0$ almost surely $\forall t\in \mathcal{D}_t\setminus \mathcal{N}_t$. One drawback of performing $p$ consecutive tests is that the initial prefixed significance level is violated if this is not modified considering the total number of partial tests. As a result, this has to be adequately corrected. Techniques such as Bonferroni's correction can be easily applied to avoid this inconvenient.\\

\subsubsection{Derivation of $\widehat{\widetilde{\mathcal{S}}}^2$}\label{subsubsec:S_int}

In this Section we prove that the variance estimator considered in (\ref{S_int}) for $\widetilde{MDD}_{n}^2(Y(t)|_{X_j(t)})= \int_{\mathcal{D}_t} MDD_{n}^2(Y(t)|_{X_j(t)})dt$ estimates correctly this quantity.\\

As we mentioned above, $\widetilde{MDD}_{n}^2(Y(t)|_{X_j(t)})$ is a $\mathcal{U}$-statistic of order four. This implies that, making use of the Hoeffding decomposition, this can be expressed as
\begin{equation*}
	\widetilde{MDD}_{n}^2(Y(t)|_{X_j(t)})= \frac{1}{\binom{n}{2}} \sum_{1\leq l<q\leq n} \widetilde{U_j(X_{lj}(t),X_{qj}(t))} \cdot \widetilde{V(Y_l(t),Y_q(t))} + (\mathcal{R}_n)_j
\end{equation*}
where $\widetilde{U_j(x,x')}=\int_{\mathcal{D}_t} \left\lbrace  \mathbb{E}\left[ K(x,X'_j(t)) \right] + \mathbb{E}\left[ K(X_j(t),x') \right] - K(x,x') - \mathbb{E}\left[ K(X_j(t),X'_j(t)) \right] \right\rbrace dt$ and $\widetilde{V(y,y')}=\int_{\mathcal{D}_t} (y-\mu_Y)(y'-\mu_Y)dt$ for $\mu_Y=\mathbb{E}[Y(t)]$, being $(\mathcal{R}_n)_j$ the remainder term.\\

Calculation about Hoeffding decomposition is quite similar to the one of Section 1.2 of the supplementary material in \cite{Zhang2018}, but taking under consideration the integrated version. The adaptation of this for our context is collected in Section \ref{append:Hoeffding} of the Appendix.\\

If we define the theoretical test statistic
\begin{equation*}
	\breve{T}_n=\sqrt{\binom{n}{2}}\frac{ \sum_{j\in D} \widetilde{MDD}_{n}^2(Y(t)|_{X_j(t)})}{\widetilde{\mathcal{S}}},
\end{equation*}
considering $\widetilde{\mathcal{S}}$ the true integrated version of the variance, we can see that
\begin{equation*}
	\begin{split}
		\breve{T}_n=& \sum_{j\in D} \frac{1}{\sqrt{\binom{n}{2}}\widetilde{\mathcal{S}}} \sum_{1\leq l<q\leq n} \widetilde{U_j(X_{lj}(t),X_{qj}(t))} \cdot \widetilde{V(Y_l(t),Y_q(t))} + \frac{\sqrt{\binom{n}{2}}}{\widetilde{\mathcal{S}}} \sum_{j\in D} (\mathcal{R}_n)_j\\	
		=& \frac{1}{\widetilde{\mathcal{S}}} (D_{n,1}+ D_{n,2})
	\end{split}
\end{equation*}
with $D_{n,1}= \binom{n}{2}^{-1/2}\sum_{1\leq l<q\leq n} \sum_{j\in D} \widetilde{U_j(X_{lj}(t),X_{qj}(t))} \cdot \widetilde{V(Y_l(t),Y_q(t))}$ is the leading term and $D_{n,2}=\binom{n}{2}^{1/2} \sum_{j\in D} (\mathcal{R}_n)_j$ is the remainder term. Under the $H_0$ assumption of (\ref{exp_contrast}) we have that
\begin{equation*}
	\Var \left[ D_{n,1} \right] = \sum_{j,j'\in D} \mathbb{E} \left[ \widetilde{V(Y(t),Y'(t))}^2 \right] \widetilde{U_j(X_{j}(t),X'_{j}(t))} \cdot \widetilde{U_{j'}(X_{j'}(t),X'_{j'}(t))}
\end{equation*}

Due to the fact that the contribution from $D_{n,2}$ is asymptotically negligible, we may set $\widetilde{\mathcal{S}}^2=\Var \left[ D_{n,1} \right]$ and then construct the variance estimator displayed in (\ref{S_int}).\\

The preliminary assumption that all curves points are observed can be quite restrictive in practice. Next, we show in Section \ref{subsec:dif_inst} how to adapt this requirement to contexts where curves are observed at different time points, adjusting the procedure to more realistic situations.

\subsubsection{Data measured at different time instants}\label{subsec:dif_inst}

\begin{figure}[htb]
	\centering
	\includegraphics[width=\textwidth]{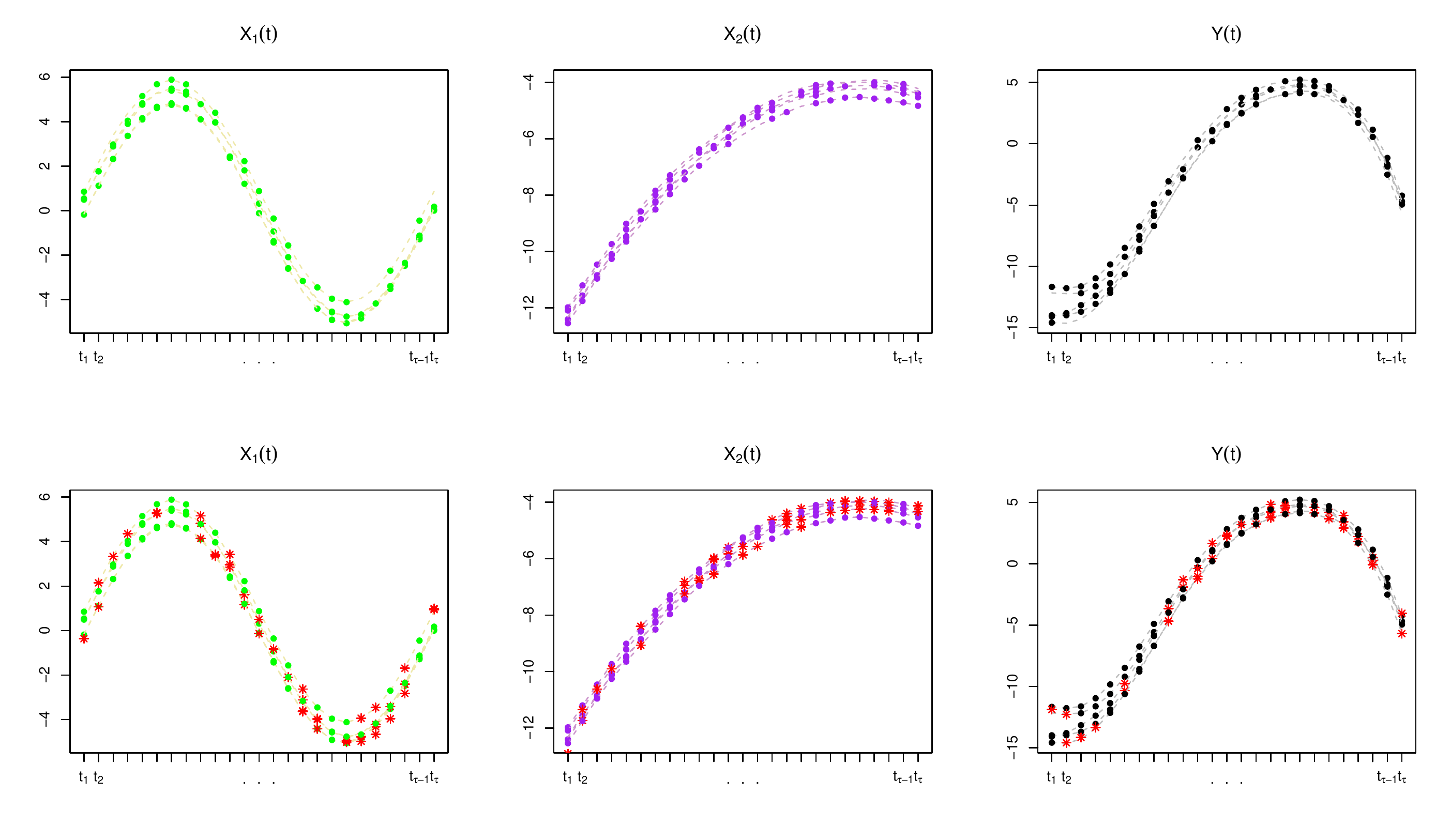}
	\caption{First row: sample of five curves measured at different time instants $\{t_u\}_{u=1}^{\mathcal{T}}\in\mathcal{D}_t$ considering $p=2$ covariates ($X_1(t)$ and $X_2(t)$) to explain $Y(t)$. Second row: same example adding the recovered points by means of splines interpolation. Filled dots ($\bullet$) represent the $n_u$ observed points at each instant $t_u$ and asterisks (\textcolor{red}{$\ast$}) the recovered nonobserved ones.}
	\label{example_param_different_obs_recov}
\end{figure}

Until now, we work under the assumption that curves were observed in practice at same time instants. In contrast, in this part we assume that a total of $n$ curves and $\{t_u\}_{u=1}^{\mathcal{T}}\in\mathcal{D}_t$ time points are considered, but we allow the observed points of each curve to be measured in a different number of instants. Then, for each time point $t_u$ there are $1\leq n_u\leq n$ observed samples of the form $\{Y_{i_u}(t_u),X_{i_u}(t_u)\}_{i_u=1}^{n_u}$. A graphic example of our current situation considering $p=2$ covariates for a concurrent model with structure similar to (\ref{multi_conc_mod}) is displayed in first row of Figure \ref{example_param_different_obs_recov}. In this example we have $n=5$ curves and, as an example, for $t_1$ there are $n_1=4$ points.\\

In this context, our proposed method can not be applied directly. This is because it is not verified $n_u=n$ for all $u=1,\dots,\mathcal{T}$. However, this problem can be solved estimating the missing curves values. This translates in a recovering of the whole curves trajectories on the grid $\{t_u\}_{u=1}^{\mathcal{T}}\in\mathcal{D}_t$, verifying now that $n_u=n$ for all $u=1,\dots,\mathcal{T}$.\\

One simple but efficient idea is to recover the complete trajectory of the curves by means of some interpolating method with enough flexibility. For example, making use of cubic spline interpolation ideas for each of the $1,\dots,n$ curves. Results of this recovery for our example are displayed in second row of Figure \ref{example_param_different_obs_recov}. In this case, we have employed the \texttt{spline} function of the \texttt{stats} library of the R software (\cite{R}).\\

In addition, other approaches to recover these points are also available. Next, we propose one based on functional basis representation. Following the guidelines of \cite{Kim2018Additive}, \cite{Ghosal2020} and \cite{Ghosal2022}, if we can assume that the total number of time observations $\stackbin[u=1]{\mathcal{T}}{\bigcup} t_u$ is dense in $\mathcal{D}_t$ then the eigenvalues and eigenfunctions corresponding
to the original curves can be estimated using functional principal component analysis (see \cite{Yao2005}). We refer to \cite{Yao2005} for more details about the procedure. As a result, we get the estimated trajectory $\hat{X}_{ij}(\cdot)$ of the true curves $X_{ij}(\cdot)$ for $i=1,\dots,n$ and $j=1,\dots,p$, given by $\hat{X}_{ij}(t)=\hat{\mu}_j(t)+\sum_{k=1}^{K} \hat{\zeta}_{ikj} \hat{\Psi}_{sj}(t)$. Here, $K$ denotes the number of considered eigenfunctions, which can be chosen by means of a prefixed percent of variance explained criterion. Consequently, we can recover the value of $Y(\cdot),X_1(\cdot),\dots,X_p(\cdot)$ on all grid $\{ t_u \}_{u=1}^\mathcal{T} \in \mathcal{D}_t$. Thus, we can work again in the context of data measure at same time instants. This procedure can be easily implemented making use of the \texttt{fpca.sc} function belonging to the library \texttt{refund} of R (see \cite{refund}). For our naive example, we have obtained similar results to the splines interpolation methodology displayed in Figure \ref{example_param_different_obs_recov}. As a result, they are omitted.


\section{Simulation studies}\label{sec:simu_scenarios}

In this Section, two simulated concurrent model scenarios are considered to assess the performance in practice of the new significance tests introduced above.  We distinguish between linear (Scenario A) and nonlinear (Scenario B) formulation of model (\ref{multi_conc_mod}). For sake of simplicity, we consider only the case where data is measure at same time instants. For this aim, a Monte Carlo study with $M=2000$ replicas on each case is carried out using the R software (\cite{R}). Besides, we compare the performance of our test with two competitors. These are the one introduced in \cite{GHOSAL2022Score}, developed in the linear framework, and the method of \cite{Kim2018Additive} for the additive formulation. Henceforth, we refer to them by FLCM and ANFCM, respectively. We refer to Section \ref{append:comp_simu} of the Appendix for more details about competitors implementation.\\

\begin{itemize}
	\item{\textbf{Scenario A (Linear model)}: We assume linearity in (\ref{multi_conc_mod}), take $t\in\mathcal{D}_t=[0,1]$ and consider $p=2$ covariates entering the model.\\
		
		As a result, the simulated model is given by the structure
		\begin{equation*}
			Y(t)=\beta_1(t)X_1(t)+\beta_2(t)X_2(t)+\varepsilon(t)
		\end{equation*}
		with
		\begin{equation*}
			X_1(t)=5\sin\left(\frac{24\pi t}{12}\right)+\varepsilon_1(t),\quad X_2(t)=\frac{-(24t-20)^2}{50}-4+\varepsilon_2(t).
		\end{equation*}
		Here, $\beta_1(t)=-\left(\frac{24t-15}{10}\right)^2-0.8$ and $\beta_2(t)=0.01((24t-12)^2 - 12^2 +100 )$. The error terms represented by $\varepsilon_1(t), \varepsilon_2(t)$ and $\varepsilon(t)$ are simulated as random gaussian processes with exponential variogram $\Omega(s,t)=0.1 e^{ \left( -\frac{24|s-t|}{10} \right) }$. We assume that a total number of $\mathcal{T}=25$ equispaced instants are observed in $\mathcal{D}_t=[0,1]$ ($\{t_u\}_{u=1}^{25}$) and there are $n=20,40,60,80,100$ curves available for each of them. An example of these functions is displayed in Figure \ref{variables_ex2_vs2}. We remark that we have not included intercept in our linear formulation because this can be done without loss of generality just centering both $Y(t)$ and $X(t)=(X_1(t),X_2(t))^\top\in\mathbb{R}^2$ for all $t\in \mathcal{D}_t$.\\
		\begin{figure}[htb]
			\centering
			\includegraphics[width=\textwidth]{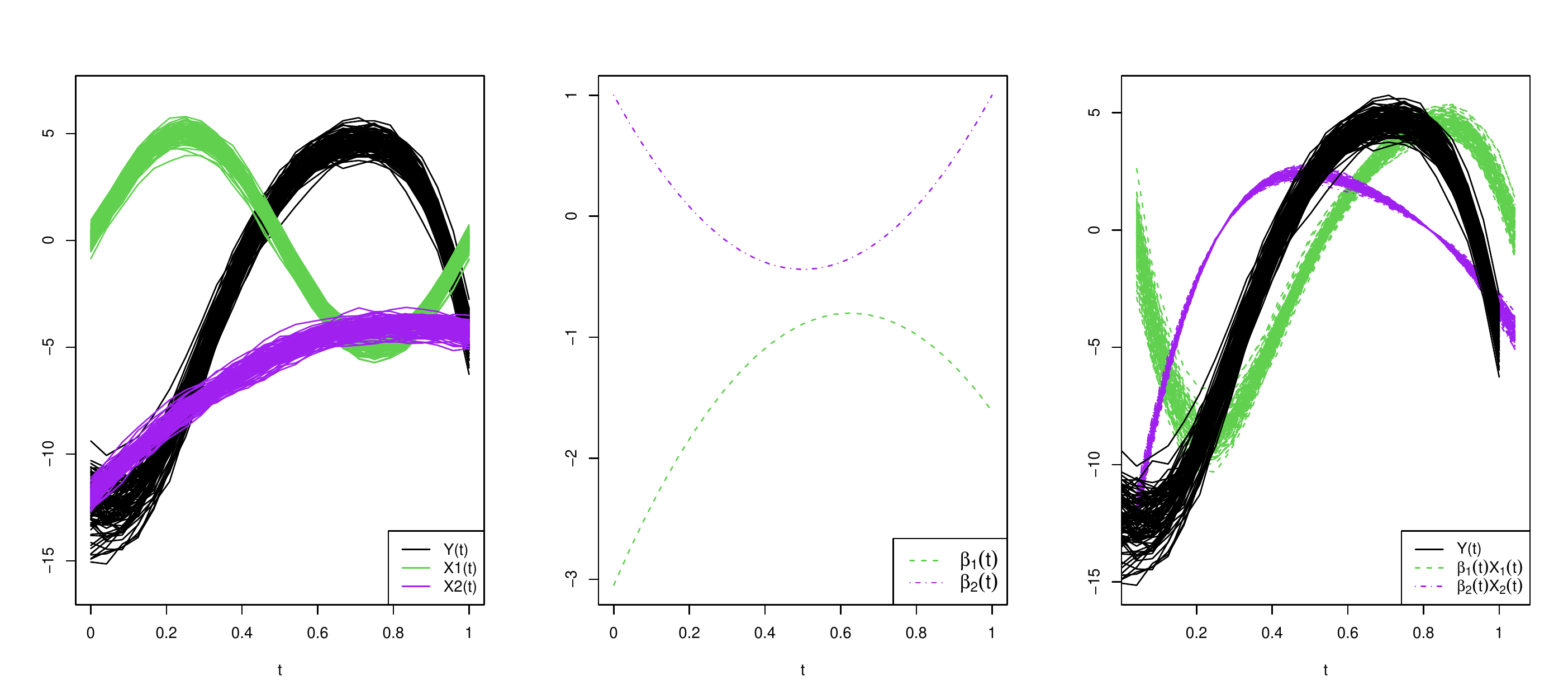}
			\caption{Left: simulated sample values of the functional variables along the grid $[0,1]$ taking $n=20$. Middle: real partial effects corresponding to $X_1(t)$ ($\beta_1(t)$) and $X_2(t)$ ($\beta_2(t)$). Right: simulated regression model components $\beta_1(t)X_1(t)$ and $\beta_2(t)X_2(t)$.}
			\label{variables_ex2_vs2}
		\end{figure} }
	\item{\textbf{ Scenario B (nonlinear model)}: a nonlinear structure of (\ref{multi_conc_mod}) is assumed for this scenario. Again, we take $t\in\mathcal{D}_t=[0,1]$ and consider $p=2$ covariates to explain the model.\\
		
	Then, this model has the expression
	\begin{equation*}
		Y(t)=F_1(t,X_1(t))+F_2(t,X_2(t))+\varepsilon(t)
	\end{equation*}
	being
	\begin{equation*}
		F_1(t,X_1(t))=\exp((24 t+1)\cdot X_1(t)/20)-2,\quad F_2(t,X_2(t))=-1.2\log(X_2(t)^2)\cdot \sin(2\pi t)
	\end{equation*}
	with $X_1(t)$ and $X_2(t)$ defined equal as in the linear case (Scenario A) and using the same observed discretization time points. Now, the errors $\varepsilon_1(t), \varepsilon_2(t)$ and $\varepsilon(t)$ are assumed to be random gaussian processes with exponential variogram $\Omega(s,t)=0.02 e^{ \left( -\frac{24|s-t|}{10} \right) }$. An example of this scenario is displayed in Figure \ref{variables_ex2_vs2_NL}.
	\begin{figure}[htb]
		\centering
		\includegraphics[width=\textwidth]{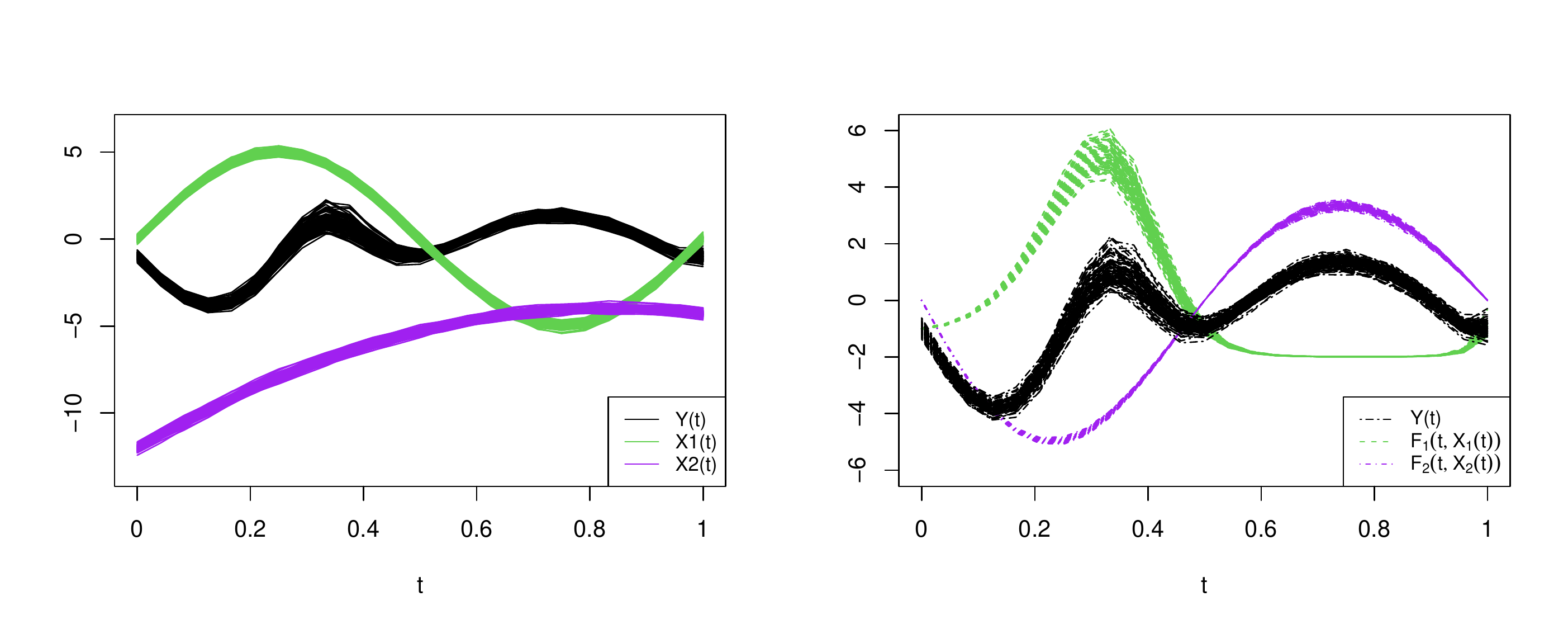}
		\caption{Left: simulated sample values of the functional variables along the grid $[0,1]$ taking $n=20$. Right: real $Y(t)$ structure jointly with partial effects corresponding to $X_1(t)$ ($F_1(t,X_1(t))$) and $X_2(t)$ ($F_2(t,X_2(t))$).}
		\label{variables_ex2_vs2_NL}
	\end{figure}
 }	
\end{itemize}

In all tests we make use of the wild bootstrap techniques introduced above in Section \ref{sec:test_MDD} to approximate the p-values. We have employed $B=1000$ resamples on each case. Besides, as we mentioned before, sample test size and power are obtained by Monte Carlo techniques. In order to know if the p-values under the null take an adequate value, we obtain the $95\%$ confidence intervals of the significance levels making use of expression $\left[ \alpha \mp 1.96 \sqrt{\frac{\alpha (1-\alpha)}{M}}  \right]$. Here $\alpha$ is the expected level and $M$ is the number of  Monte Carlo simulated samples. As a result, we consider that a p-value is acceptable for levels $\alpha=0.01,0.05,0.1$ if this is within the values collected in Table \ref{CI_p_values} for the Monte Carlo replicates. We highlight the values out of these scales in \textbf{bold} for simulation results.\\

\begin{table}[htb] 
	\centering
	\small 
	\begin{tabular}{c ccc }
		\toprule
		\textbf{M} & $\boldsymbol{\alpha=0.01}$ & $\boldsymbol{\alpha=0.05}$ & $\boldsymbol{\alpha=0.1}$ \\
		\hline
		\multicolumn{1}{c}{\rule{0pt}{0.4cm} $1000$}  & $[0.004, 0.016]$ & $[ 0.036, 0.064]$ & $[0.081, 0.119]$ \\
		\multicolumn{1}{c}{\rule{0pt}{0.4cm} $2000$}  & $[0.006, 0.014]$ & $[0.040, 0.060]$ & $[0.087, 0.113]$ \\
		\bottomrule
	\end{tabular}
	\caption{Confidence intervals at $95\%$ of the Monte Carlo proportions for $M$ replicates. }
	\label{CI_p_values}
\end{table}

\subsection{Results for scenario A (linear model)}\label{subsec:simu_linear}

We start analyzing the performance of the global mean dependence test in the linear model formulation, making use of Scenario A introduce above in Section \ref{sec:simu_scenarios}. For this purpose, we consider three different scenarios. In the first one, the null hypothesis of mean independence is verified simulating under the assumption that $\beta_1(t)=\beta_2(t)=0$. Next, we consider the remaining two  cases under the alternative that information provided by $X(t)=(X_1(t),X_2(t))^\top$ is useful in some way: only the $X_2(t)$ covariate is relevant (fixing $\beta_1(t)=0$) or both covariates $X_1(t)$ and $X_2(t)$ support relevant information to correctly explain $Y(t)$.\\

\begin{table}[htb] 
	\centering
	\small 
	\begin{tabular}{c ccc ccc ccc}
		\toprule
		\textbf{Model:} & \multicolumn{3}{c}{$\boldsymbol{\beta_1(t)=\beta_2(t)=0}\; ( H_0 )$} & \multicolumn{3}{c}{$\boldsymbol{\beta_1(t)=0, \beta_2(t)\not=0}\; ( H_a )$} & \multicolumn{3}{c}{$\boldsymbol{\beta_1(t)\not=0, \beta_2(t)\not=0}\; ( H_a )$}\\
		\cmidrule(r){2-4} \cmidrule(rl){5-7} \cmidrule(l){8-10}
		& \rule{0pt}{0.4cm} \textbf{1\%} & \textbf{5\%} & \textbf{10\%} & \textbf{1\%} & \textbf{5\%} & \textbf{10\%} & \textbf{1\%} & \textbf{5\%} & \textbf{10\%} \\
		\hline
		\multicolumn{1}{c}{\rule{0pt}{0.4cm} $n=20$} & 0.010 & 0.045 & 0.092 & 0.574 & 0.797 & 0.882 & 1 & 1 & 1 \\
		\multicolumn{1}{c}{$n=40$}  & 0.013 & 0.050 & 0.093 & 0.984 & 0.998 & 1 & 1 & 1 & 1  \\
		\multicolumn{1}{c}{$n=60$}  & 0.007 & 0.052 & 0.103 & 1 & 1 & 1 & 1 & 1 & 1 \\  
		\multicolumn{1}{c}{$n=80$}  & 0.009 & 0.045 & 0.094 & 1 & 1 & 1 & 1 & 1 & 1 \\
		\multicolumn{1}{c}{$n=100$} & 0.012 & 0.050 & 0.088 & 1 & 1 & 1 & 1 & 1 & 1 \\
		\bottomrule
	\end{tabular}
	\caption{Empirical sizes and powers of the MDD-based global test for mean independence testing using wild bootstrap approximation with $B=1000$ resamples in Scenario A. }
	\label{Results_effect_global_boot_int}
\end{table}

Obtained results are collected in Table \ref{Results_effect_global_boot_int} for $n=20,40,60,80,100$. In view of the results, we can appreciate as the empirical sizes approximate the fixed significance levels as $n$ increases when $H_0$ is true ($H_0 : \beta_1(t)=\beta_2(t)=0$). Besides, the empirical distribution of the p-values seems to be a $U[0,1]$ as it is appreciated in Figure \ref{p_values_dist_MDD_CM_global_boot_U}. In contrast, simulating under alternative hypothesis, $H_a :\beta_1(t)=0,\beta_2(t)\not=0$ and $H_a: \beta_1(\mathfrak{t})\not=0, \beta_2(\mathfrak{t})\not=0$ scenarios, the test power tends to one as the sample size increases. As a result, we can claim that the test is well calibrated and has power.\\

Once we have rejected the null hypothesis that all covariates are irrelevant in practice, we can detect which of them play a role in terms of data explanation. For this aim, partial contrasts can be carried out, testing if every covariate is irrelevant, $H_{0j}:\beta_j(t)=0 \; \forall t \in \mathcal{D}_t$, or not, $H_{aj}:\beta_j(t)\not=0 \; \text{for some } t \in \mathcal{V}_t$, being $j=1,\dots,p$.\\

Again, we have considered different scenarios. First of all we assume that $X(t)$ is not significant taking $\beta_1(t)=\beta_2(t)=0$. Then, we move to the situation where only $X_2(t)$ is relevant and finally we consider the model including both $X_1(t)$ and $X_2(t)$ effects to explain $Y(t)$. Results for these considered simulation scenarios are displayed in Table \ref{Results_effect_global_boot_partial_U}. Here, we appreciate as the empirical sizes tend to the significance levels simulating under the null hypothesis that both covariates, separately, have not got a relevant effect on the response. Besides, we see as in case of having $\beta_1(t)=0$ and $\beta_2(t)\not=0$, these tests help us to select relevant information $X_2(t)$ and discard noisy one $X_1(t)$. Otherwise, when both covariates are relevant, the partial tests clearly reject the null hypothesis of null effect tending its powers to the unit as sample sizes increase.\\

\begin{table}[htb] 
	\centering
	\small
	\begin{tabular}{c cc cc cc}
		\toprule
		\textbf{Model:}& \multicolumn{2}{c}{$\boldsymbol{\beta_1(t)=\beta_2(t)=0}$} & \multicolumn{2}{c}{$\boldsymbol{\beta_1(t)=0,\; \beta_2(t)\not=0}$} & \multicolumn{2}{c}{$\boldsymbol{\beta_1(t)\not=0,\; \beta_2(t)\not=0}$} \rule{0pt}{0.25cm}\\
		\cmidrule(r){1-3} \cmidrule(rl){4-5} \cmidrule(l){6-7}
		& \multicolumn{1}{c}{\small $H_{01}$} & \multicolumn{1}{c}{\small $H_{02}$} & \multicolumn{1}{c}{\small $H_{01}$} & \multicolumn{1}{c}{\small $H_{02}$} & \multicolumn{1}{c}{\small $H_{01}$} &  \multicolumn{1}{c}{\small $H_{02}$} \\ 
		\cmidrule(r){2-2} \cmidrule(rl){3-3} \cmidrule(rl){4-4} \cmidrule(rl){5-5} \cmidrule(rl){6-6} \cmidrule(l){7-7}
		& \rule{0pt}{0.4cm} \textbf{5\%}/\textbf{10\%} & \textbf{5\%}/\textbf{10\%} & \textbf{5\%}/\textbf{10\%} & \textbf{5\%}/\textbf{10\%} & \textbf{5\%}/\textbf{10\%} & \textbf{5\%}/\textbf{10\%} \\
		\hline
		\multicolumn{1}{c}{\rule{0pt}{0.4cm} $n=20$} & 0.040/\textbf{0.078} & 0.043/0.101 & 0.041/0.087 & 0.919/0.966 & 1/1 & 0.330/0.490 \\
		\multicolumn{1}{c}{$n=60$}  & 0.048/0.101 & 0.049/0.103 & 0.047/0.098 & 1/1 & 1/1 & 0.935/0.971 \\  
		\multicolumn{1}{c}{$n=100$} & 0.046/0.089 & 0.047/0.096 & 0.046/\textbf{0.086} & 1/1 & 1/1 & 0.998/1 \\
		\bottomrule
	\end{tabular}
	\caption{Empirical sizes and powers of the partial MDD-based global tests for mean independence testing considering $H_{01}: \mathbb{E}[Y(t)|_{X_1(t)}]=\mathbb{E}[Y(t)]$ and $H_{02}: \mathbb{E}[Y(t)|_{X_2(t)}]=\mathbb{E}[Y(t)]$ and using wild bootstrap approximation with $B=1000$ resamples in Scenario A. }
	\label{Results_effect_global_boot_partial_U}
\end{table}

\subsection{Results for scenario B (nonlinear model)}\label{subsec:simu_non_linear}

In this Section the performance of the MDD global mean independence test is analyzed in a more difficult framework: a nonlinear effects formulation. For this aim, we make use of Scenario B introduced in Section \ref{sec:simu_scenarios}. Again, we consider three different situations of dependence, following the same arguments of Section \ref{subsec:simu_linear}. As a result, we simulate a situation of no effect ($H_0:F_1(t, X_1(t))=F_2(t, X_2(t))=0$) which corresponds with independence and two dependence frameworks where only one covariate is relevant ($H_a:F_1(t, X_1(t))=0, F_2(t, X_2(t))\not=0$) or both of them are ($H_a:F_1(t, X_1(t))\not=0, F_2(t, X_2(t))\not=0$).\\

\begin{table}[htb] 
	\centering
	\small 
	\begin{tabular}{c ccc ccc ccc}
		\toprule
		\textbf{Model:} & \multicolumn{3}{c}{$\boldsymbol{F_1(\cdot)=F_2(\cdot)=0}\; ( H_0 )$} & \multicolumn{3}{c}{$\boldsymbol{F_1(\cdot)=0, F_2(\cdot)\not=0}\; ( H_a )$} & \multicolumn{3}{c}{$\boldsymbol{F_1(\cdot)\not=0, F_2(\cdot)\not=0}\; ( H_a )$}\\
		\cmidrule(r){2-4} \cmidrule(rl){5-7} \cmidrule(l){8-10}
		& \rule{0pt}{0.4cm} \textbf{1\%} & \textbf{5\%} & \textbf{10\%} & \textbf{1\%} & \textbf{5\%} & \textbf{10\%} & \textbf{1\%} & \textbf{5\%} & \textbf{10\%} \\
		\hline
		\multicolumn{1}{c}{\rule{0pt}{0.4cm} $n=20$} & 0.011 & 0.049 & 0.096 &  0.215 & 0.426 & 0.563  &  0.989 & 1 & 1 \\
		\multicolumn{1}{c}{$n=40$}  & 0.013 & 0.05 & 0.094 & 0.564 & 0.793 & 0.886 & 1 & 1 & 1 \\
		\multicolumn{1}{c}{$n=60$}  & 0.009 & 0.053 & 0.105 & 0.871 & 0.956 & 0.979 & 1 & 1 & 1 \\  
		\multicolumn{1}{c}{$n=80$}  & 0.01 & 0.046 & 0.096 & 0.974 & 0.996 & 1 & 1 & 1 & 1 \\
		\multicolumn{1}{c}{$n=100$} & 0.013 & 0.054 & 0.093 & 0.994 & 1 & 1 & 1 & 1 & 1 \\
		\bottomrule
	\end{tabular}
	\caption{Empirical sizes and powers of the MDD-based global test for mean independence testing using wild bootstrap approximation with $B=1000$ resamples in Scenario B. }
	\label{Results_effect_global_boot_int_NL}
\end{table}

Results of the $M=2000$ Monte Carlo simulations taking $n=20,40,60,80,100$ are displayed in Table \ref{Results_effect_global_boot_int_NL} for the MDD-test. We can appreciate as in case of simulating under the null hypothesis $H_0$ the p-values tend to stabilize around the significance levels. In fact, Figure \ref{p_values_dist_MDD_CM_global_boot_NL} shows as these seem to follow a uniform distribution in $[0,1]$. Then, we can conclude that our test is well-calibrated even for nonlinear approaches. In terms of the power, we can see as when the independence assumption is violated the p-values tend to $1$ as sample size increased. Two examples of this phenomenon are displayed in Table \ref{Results_effect_global_boot_int_NL} simulating under different alternative hypothesis. Summing up, we have a well-calibrated and powerful test  in a nonlinear framework.\\

Next, our interest focus on partial tests to be able to apply covariates selection in this nonlinear scenario Again, we consider the three different dependence scenarios introduced above but testing the independence for each covariate separately, applying a total of $j=1,\dots,p$ tests. In this way, we expect for situation as $F_1(t, X_1(t))=0$, $F_2(t, X_2(t))\not=0$ to be capable of detecting relevant covariates ($X_2(t)$), rejecting its corresponding $H_{0j}$ hypothesis, and to exclude noise ones from the model otherwise ($X_1(t)$). Results for partial tests are collected in Table \ref{Results_effect_global_boot_partial_U_NL}. We can see as these tests allow us to determine which covariates play a relevant role on each scenario, with p-values greater than the significance levels and tending to $1$ as sample size increases, and classify as irrelevant those verifying that its associated p-values are less or equal than significance levels.\\

\begin{table}[htb] 
	\centering
	\small
	\begin{tabular}{c cc cc cc}
		\toprule
		\textbf{Model:}& \multicolumn{2}{c}{$\boldsymbol{F_1(\cdot)=F_2(\cdot)=0}$} & \multicolumn{2}{c}{$\boldsymbol{F_1(\cdot)=0,\; F_2(\cdot)\not=0}$} & \multicolumn{2}{c}{$\boldsymbol{F_1(\cdot)\not=0,\; F_2(\cdot)\not=0}$} \rule{0pt}{0.25cm}\\
		\cmidrule(r){1-3} \cmidrule(rl){4-5} \cmidrule(l){6-7}
		& \multicolumn{1}{c}{\small $H_{01}$} & \multicolumn{1}{c}{\small $H_{02}$} & \multicolumn{1}{c}{\small $H_{01}$} & \multicolumn{1}{c}{\small $H_{02}$} & \multicolumn{1}{c}{\small $H_{01}$} &  \multicolumn{1}{c}{\small $H_{02}$} \\ 
		\cmidrule(r){2-2} \cmidrule(rl){3-3} \cmidrule(rl){4-4} \cmidrule(rl){5-5} \cmidrule(rl){6-6} \cmidrule(l){7-7}
		& \rule{0pt}{0.4cm} \textbf{5\%}/\textbf{10\%} & \textbf{5\%}/\textbf{10\%} & \textbf{5\%}/\textbf{10\%} & \textbf{5\%}/\textbf{10\%} & \textbf{5\%}/\textbf{10\%} & \textbf{5\%}/\textbf{10\%} \\
		\hline
		\multicolumn{1}{c}{\rule{0pt}{0.4cm} $n=20$} & 0.04/\textbf{0.078}  & 0.043/0.101  & 0.04/\textbf{0.077}  & 0.567/0.692 & 1/1 & 0.18/0.299 \\
		\multicolumn{1}{c}{$n=60$}  & 0.048/0.101  & 0.049/0.103 & 0.053/0.107  & 0.987/0.995 & 1/1 & 0.621/0.783  \\ 
		\multicolumn{1}{c}{$n=100$} & 0.046/0.089 & 0.047/0.096 & 0.044/0.09  & 1/1 & 1/1 & 0.915/0.971  \\ 
		\bottomrule
	\end{tabular}
	\caption{Empirical sizes and powers of the partial MDD-based global tests for mean independence testing considering $H_{01}: \mathbb{E}[Y(t)|_{X_1(t)}]=\mathbb{E}[Y(t)]$ and $H_{02}: \mathbb{E}[Y(t)|_{X_2(t)}]=\mathbb{E}[Y(t)]$ and using wild bootstrap approximation with $B=1000$ resamples in Scenario B. }
	\label{Results_effect_global_boot_partial_U_NL}
\end{table}

\subsection{Comparison with FLCM and ANFCM algorithms}\label{subsec:simu_comaprison}

Next, we want to compare our novel procedure with existing competitors in literature. For this aim, we have considered the FLCM algorithm of \cite{GHOSAL2022Score} for the linear framework and the ANFCM procedure of \cite{Kim2018Additive} for a more flexible model assuming additive effects. Both of them have displayed really good results in practice for a proper selection of the tuning parameters for linear and additive effects, respectively.\\

In our simulation scenarios we have considered a dependence structure where all instants are related between them, emulating a real functional dataset. Nevertheless, in \cite{GHOSAL2022Score} and \cite{Kim2018Additive} this does not apply in their simulation scenarios. As a result, to perform a fair competition, we start analyzing the behavior of our MDD-based tests comparing this with results of FLCM in Scenario A of \cite{GHOSAL2022Score} and results of ANFCM for Scenario (B) taking error $\text{E}^3$ introduced in the testing performance study of \cite{Kim2018Additive}. In this last, a modification is considered to perform Algorithm 1, considering only the second covariate with nonlinear effect. In both cases, we simulate under the dense assumption being $\{t_u\}_{u=1}^{81}$ a total of $m=81$ equidistant time points in $[0,1]$. We  keep the authors parameters selection and perform a Monte Carlo study with $M=1000$ samples in all cases, obtaining the p-values for means of $B=200$ bootstrap replications. Besides, following author's recommendation after a preliminary study to determine the optimal number of basis functions for these examples, we work with 7 components for both, FLCM and ANFCM procedures. More details can be found in \cite{GHOSAL2022Score} or \cite{Kim2018Additive}, respectively. We remind the scenarios structure and explain implementation issues in Section \ref{append:comp_simu} of the Appendix.\\

Results of the comparison between FLCM and MDD effect tests for scenario A of \cite{GHOSAL2022Score} are showed in Table \ref{Results_effect_NL_FLCM_MDD}. We can appreciate that, simulating under the null ($d=0$), one value of the FLCM algorithm are out of the $95\%$ p-values confidence intervals for $M=1000$ whereas the MDD does not suffer from this issue. Moreover, paying attention to the p-values distributions under the null, which are displayed in Figure \ref{p_values_comparison_FLCM_MDD.pdf}, we can see as the FLCM p-values do not follow a uniform distribution. In contrast, this is corrected making use of our MDD based test. As a result, it seems that our test provides a bit better calibration than the FLCM. In terms of the power, levels for both algorithms tends to 1 as sample size increases and their values are greater for the $d=7$ scenario than the $d=3$ one, as it would be expected. Now, FLCM algorithm outperforms the MDD results in all scenarios. However, our procedure is still quite competitive even taking under consideration that the data is simulating under the linear assumption, given advantage to the FLCM procedure.\\

\begin{table}[htb] 
	\centering
	\small
	\begin{tabular}{c c ccc ccc ccc}
		\toprule
		\multicolumn{2}{c}{\textbf{Model:}} &  \multicolumn{3}{c}{$\mathbf{H_0\, (d=0)}$} & \multicolumn{3}{c}{$\mathbf{H_a \, (d=3)}$} & \multicolumn{3}{c}{$\mathbf{H_a \, (d=7)}$} \rule{0pt}{0.25cm}\\
		\cmidrule(r){3-5} \cmidrule(rl){6-8} \cmidrule(l){9-11}
		& \rule{0pt}{0.4cm} & \textbf{1\%} & \textbf{5\%} & \textbf{10\%} & \textbf{1\%} & \textbf{5\%} & \textbf{10\%} & \textbf{1\%} & \textbf{5\%} & \textbf{10\%} \\
		\hline
				
		\multirow{2}{*}{n=60} &\rule{0pt}{0.4cm} FLCM & 0.007 & 0.054 & 0.103 & 0.776 & 0.888 & 0.937 & 0.999 & 1 & 1 \\
		& MDD & 0.014 & 0.052 & 0.097 & 0.341 & 0.550 & 0.671 & 0.992 & 0.997 & 1 \\		
				
		\multirow{2}{*}{n=100}&\rule{0pt}{0.4cm} FLCM & 0.005 & 0.038 & \textbf{0.077} & 0.964 & 0.979 & 0.992 & 1 & 1 & 1 \\
		 & MDD & 0.013 & 0.049 & 0.103 & 0.619 & 0.796 & 0.871 & 1 & 1 & 1 \\
		\bottomrule
	\end{tabular}
	\caption{Summary of empirical sizes and powers of the FLCM and MDD effect tests. }
	\label{Results_effect_NL_FLCM_MDD}
\end{table}

Next, we compare the performance of the MDD with the ANFCM approach in an additive framework. Simulation results are collected in Table \ref{Results_effect_NL_ANFCM_MDD} for both procedures. We see that both methodologies are well calibrated under the null ($d=0$) for all levels except for the $1\%$, where their values are out of the $95\%$ p-values confidence interval for $n=60$. Nevertheless, this issue is solved for greater values of $n$, as it is exemplified for $n=100$. Moreover, simulating under $H_0$, the p-values follow a uniform distribution as it can be appreciated in Figure \ref{p_values_comparison_ANFCM_MDD.pdf}. If we simulate under alternative hypothesis considering an additive effect ($d=3$ and $d=7$) we see as these quantities tend to $1$ as sample size increases. In addition, as the covariate effect becomes more noticeable, going from $d=3$ to $d=7$, the power of ANFCM and MDD procedures increases. Again, the power of the ANFCM algorithm is always greater than the MDD one. In this point, it is important to remark that the ANFCM algorithm takes advantage of the fact that an additive structure with intercept function is assumed preliminarily. In contrast, our MDD test does not assume any structure, not even the inclusion of intercept to the model, so this has to measure all possible forms of departure from conditional mean independence. \\

\begin{table}[htb] 
	\centering
	\small
	\begin{tabular}{c c ccc ccc ccc}
		\toprule
		\multicolumn{2}{c}{\textbf{Model:}} &  \multicolumn{3}{c}{$\mathbf{H_0\, (d=0)}$} & \multicolumn{3}{c}{$\mathbf{H_a \, (d=3)}$} & \multicolumn{3}{c}{$\mathbf{H_a \, (d=7)}$} \rule{0pt}{0.25cm}\\
		\cmidrule(r){3-5} \cmidrule(rl){6-8} \cmidrule(l){9-11}
		& \rule{0pt}{0.4cm} & \textbf{1\%} & \textbf{5\%} & \textbf{10\%} & \textbf{1\%} & \textbf{5\%} & \textbf{10\%} & \textbf{1\%} & \textbf{5\%} & \textbf{10\%} \\
		\hline
		
		\multirow{2}{*}{n=60} &\rule{0pt}{0.4cm} ANFCM & \textbf{0.021} & 0.063 & 0.117 & 1 & 1 & 1 & 1 & 1 & 1 \\
		& MDD & \textbf{0.019} & 0.058 & 0.102 & 0.410 & 0.811 & 0.944 & 0.747 & 0.984 & 1 \\		
		
		\multirow{2}{*}{n=100}&\rule{0pt}{0.4cm} ANFCM & 0.014 & 0.056 & 0.094 & 1 & 1 & 1 & 1 & 1 & 1 \\
		& MDD & 0.008 & 0.046 & 0.095 & 0.929 & 0.999 & 1  & 0.999 & 1 & 1 \\
		\bottomrule
	\end{tabular}
	\caption{Summary of empirical sizes and powers of the ANFCM and MDD effect tests. }
	\label{Results_effect_NL_ANFCM_MDD}
\end{table}

It is relevant to notice that, in both previous scenarios, covariates are related with response by means of trigonometric functions when it corresponds. Then, the effects modeling takes advantage of the B-spline basis representation. This, together with the fact that errors are assumed to be independent between them for different instants, is a clear advantage for the FLCM and the ANFCM algorithms in comparison with our procedure. Thus, to test the FLCM and ANFCM performance in a functional context with time correlated errors and when the model structure does not depend on only trigonometric functions, we apply these to our simulation scenarios introduced in Section \ref{sec:simu_scenarios}. For this purpose, we consider a partial approach testing the covariates effect separately using the FLCM procedure in Scenario A and the ANFCM one in Scenario B. In order to compare our results with theirs, we simulate now $M=2000$ Monte Carlo replications and use $B=1000$ bootstraps resamples for ANFCM. Again, we follows authors recommendation and use $K=7$ basis terms in both procedures \footnote{In this setup, we have $\mathcal{T}=25$ time instants. Then, for the ANFCM procedure, as the function \texttt{fpca.face} employs by default a number of $35$ knots to carry out FPCA, we have to reduce this. We decided to take $12$ knots to solve this issue.}. We refer to Section \ref{append:comp_simu} of the Appendix for a summary of the simulation parameters selection.\\

Results of partial FLCM tests in scenario A are displayed in Table \ref{Results_effect_global_FLCM_partial}. We can appreciate as, it does not matter the sample size employed, the test is bad calibrated. In fact, all obtained p-values are out of the $95\%$ confidence intervals. This contrasts with the MDD results displayed in Table \ref{Results_effect_global_boot_partial_U}, where the test is well calibrated. This may be because, as mentioned before, a different dependence structure is considered more related with functional nature. In terms of power, there is not a clear winner, because our test is more powerful in the $H_a: \beta_1(t)=0,\; \beta_2(t)\not=0$ scenario in test $H_{02}$, but FLCM is a bit more powerful in last scenario for $H_{02}$. However, this difference is small and considering that the FLCM is not well calibrated, we can conclude that our procedure outperforms this.\\

\begin{table}[htb] 
	\centering
	\small
	\begin{tabular}{c cc cc cc}
		\toprule
		\textbf{Model:}& \multicolumn{2}{c}{$\boldsymbol{\beta_1(t)=\beta_2(t)=0}$} & \multicolumn{2}{c}{$\boldsymbol{\beta_1(t)=0,\; \beta_2(t)\not=0}$} & \multicolumn{2}{c}{$\boldsymbol{\beta_1(t)\not=0,\; \beta_2(t)\not=0}$} \rule{0pt}{0.25cm}\\
		\cmidrule(r){1-3} \cmidrule(rl){4-5} \cmidrule(l){6-7}
		& \multicolumn{1}{c}{\small $H_{01}$} & \multicolumn{1}{c}{\small $H_{02}$} & \multicolumn{1}{c}{\small $H_{01}$} & \multicolumn{1}{c}{\small $H_{02}$} & \multicolumn{1}{c}{\small $H_{01}$} &  \multicolumn{1}{c}{\small $H_{02}$} \\ 
		\cmidrule(r){2-2} \cmidrule(rl){3-3} \cmidrule(rl){4-4} \cmidrule(rl){5-5} \cmidrule(rl){6-6} \cmidrule(l){7-7}
		& \rule{0pt}{0.4cm} \textbf{5\%}/\textbf{10\%} & \textbf{5\%}/\textbf{10\%} & \textbf{5\%}/\textbf{10\%} & \textbf{5\%}/\textbf{10\%} & \textbf{5\%}/\textbf{10\%} & \textbf{5\%}/\textbf{10\%} \\
		\hline
		\multicolumn{1}{c}{\rule{0pt}{0.4cm} $n=20$} & \textbf{0.1}/\textbf{0.174} & \textbf{0}/\textbf{0.002} & \textbf{0.104}/\textbf{0.172} & 0.622/0.758 & 1/1 & 1/1 \\
		\multicolumn{1}{c}{$n=60$} & \textbf{0.09}/\textbf{0.152} & \textbf{0}/\textbf{0.004} & \textbf{0.09}/\textbf{0.158} & 0.709/0.875 & 1/1 & 1/1 \\  
		\multicolumn{1}{c}{$n=100$} & \textbf{0.074}/\textbf{0.125} & \textbf{0}/\textbf{0.003} & \textbf{0.1}/\textbf{0.17} & 0.913/0.983 & 1/1 & 1/1 \\
		\bottomrule
	\end{tabular}
	\caption{Empirical sizes and powers of the FLCM effect test considering $H_{01}: \beta_1(t)=0$ and $H_{02}: \beta_2(t)=0$ in Scenario A. }
	\label{Results_effect_global_FLCM_partial}
\end{table}

Next, performance of ANFCM algorithm is tested simulating under the Scenario B of Section \ref{sec:simu_scenarios}. Results are collected in Table \ref{Results_effect_global_ANFCM_partial_NL}. Again, comparing the ANFCM results with the ones of the MDD test (Table \ref{Results_effect_global_boot_partial_U_NL}) we see that our test is well calibrated even for small values as $n=20$ (except for a couple of exceptions) in contrast with the ANFCM one. In this last, most values are out of the $95\%$ confidence intervals. Moreover, the MDD test has more power than ANFCM in almost all cases. As a result, we can conclude that the MDD outperforms the ANFCM procedure.\\

\begin{table}[htb] 
	\centering
	\small
	\begin{tabular}{c cc cc cc}
		\toprule
		\textbf{Model:}& \multicolumn{2}{c}{$\boldsymbol{F_1(\cdot)=F_2(\cdot)=0}$} & \multicolumn{2}{c}{$\boldsymbol{F_1(\cdot)=0,\; F_2(\cdot)\not=0}$} & \multicolumn{2}{c}{$\boldsymbol{F_1(\cdot)\not=0,\; F_2(\cdot)\not=0}$} \rule{0pt}{0.25cm}\\
		\cmidrule(r){1-3} \cmidrule(rl){4-5} \cmidrule(l){6-7}
		& \multicolumn{1}{c}{\small $H_{01}$} & \multicolumn{1}{c}{\small $H_{02}$} & \multicolumn{1}{c}{\small $H_{01}$} & \multicolumn{1}{c}{\small $H_{02}$} & \multicolumn{1}{c}{\small $H_{01}$} &  \multicolumn{1}{c}{\small $H_{02}$} \\ 
		\cmidrule(r){2-2} \cmidrule(rl){3-3} \cmidrule(rl){4-4} \cmidrule(rl){5-5} \cmidrule(rl){6-6} \cmidrule(l){7-7}
		& \rule{0pt}{0.4cm} \textbf{5\%}/\textbf{10\%} & \textbf{5\%}/\textbf{10\%} & \textbf{5\%}/\textbf{10\%} & \textbf{5\%}/\textbf{10\%} & \textbf{5\%}/\textbf{10\%} & \textbf{5\%}/\textbf{10\%} \\
		\hline
		\multicolumn{1}{c}{\rule{0pt}{0.4cm} $n=20$} & \textbf{0.098}/\textbf{0.17} &\textbf{0.102}/\textbf{0.176} & \textbf{0.134}/\textbf{0.203} & 0.043/0.08 & 0.662/0.8 & 0.123/0.191  \\ 
		\multicolumn{1}{c}{$n=60$} & \textbf{0.068}/\textbf{0.118} & 0.058/0.113 & \textbf{0.071}/\textbf{0.128} & 0.013/0.028 & 1/1 & 0.117/0.216 \\  
		\multicolumn{1}{c}{$n=100$} & 0.047/0.102 & 0.055/0.106 & 0.049/0.1 & 0.063/0.114 & 1/1 & 0.147/0.293  \\ 
		\bottomrule
	\end{tabular}
	\caption{Empirical sizes and powers of the ANFCM effect test considering $H_{01}: F_1\left( t,X_1(t) \right)=0$ and $H_{02}:  F_2\left( t,X_2(t) \right)=0$ and using $B=1000$ bootstrap resamples in Scenario B. }
	\label{Results_effect_global_ANFCM_partial_NL}
\end{table}

Summing up, we have proved that our algorithm performs quite well in scenarios where the FLCM and the ANFCM procedures have an advantage, considering uncorrelated errors and trigonometric functions. Moreover, our test outperforms them when the context becomes more difficult, as in our proposed scenarios A and B where related errors and other types of relation different from trigonometric functions are considered.


\section{Real data analysis}\label{sec:real_data}

Here we test the performance of the proposed algorithms in three real data sets. Firstly, the well-known gait dataset of \cite{Gait1989} is considered as an example of linear effects model with data measure at same time instants. This has been already studied in the concurrent model framework in works as the one of \cite{GHOSAL2022Score} or \cite{Kim2018Additive}. Next, a google flu database from the USA, borrowed from \cite{wang2017unified}, is considered. In this work, a linear concurrent model formulation is assumed to model the data. Eventually, we have considered an example of nonlinear and measured at different time instants model. For this aim, the bike sharing dataset of \cite{Fanaee2014} is analyzed. We compare our results with the ones of \cite{Ghosal2022} in this concurrent model framework.

\subsection{Gait data}

In this section we analyze the new dependence test performance with a well-known dataset from the functional data context. This is the gait database (\cite{Gait1989}, \cite{ramsay2004functional}) in which the objective is to understand how the joints in hip and knee interact during a gait cycle in children. This problem has been already studied in the concurrent model context with different methodology (see \cite{GHOSAL2022Score} or \cite{Kim2018Additive}). As a consequence, we compare our results with theirs.\\

\begin{figure}[htb]
	\centering
	\includegraphics[width=\textwidth]{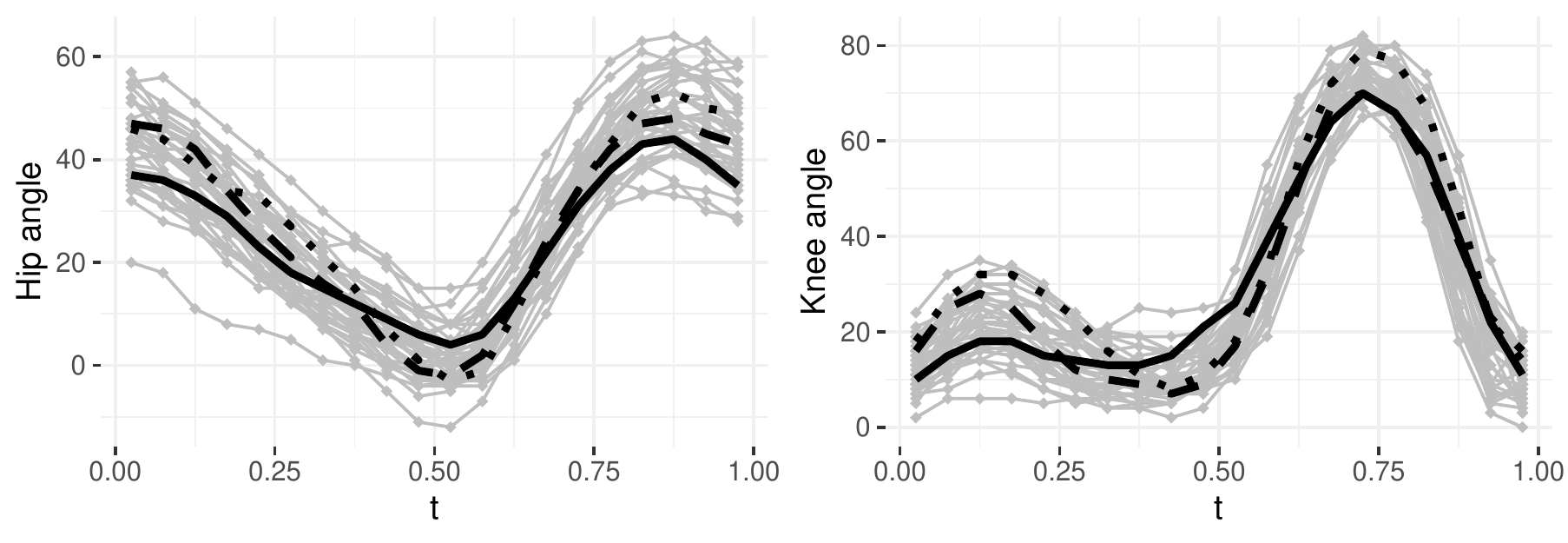}
	\caption{Hip (left) and knee (right) angles measurements of a complete gait cycle. }
	\label{data_gait_plot}
\end{figure}

The data consist of longitudinal measurements of hip and knee angles taken on 39 children with gait deficiency as they walk through a single gait cycle. This data can be found in the \texttt{fda} library (\cite{Ramsay2020}) of the R software (\cite{R}). The hip and knee angles are measured at 20 evaluation points $\{t_u\}_{u=1}^{20}$ in $[0,1]$, which are translated from percent values of the cycle. Following previous studies, we have considered as response $Y(t)$ the knee angle and as explanatory covariate $X(t)$ the hip angle. Data is displayed in Figure \ref{data_gait_plot}.\\

Applying our dependence test we obtain a p-value$\approx0$. Thus, we have strong enough evidence to reject the independence hypothesis to the usual significance levels. This translates in that there exists dependence between knee and hip angle in a gait data cycle in children with deficient gait. This result agrees with the ones of \cite{Kim2018Additive} or \cite{GHOSAL2022Score} among others in the concurrent model framework. They obtain p-values less than $0.004$ and $0.001$, respectively. Summing up, the hip angle measured at a specific time point in a gait cycle has a strong effect on the knee angle at the same time point in children.\\

\subsection{Google flu data from U.S.A}

Google flu data is used in \cite{wang2017unified} to model the relationship between flu activity and temperature fluctuation in the USA. For this purpose, influenza-like illness (ILI) cases per 100000 doctor visits are considered in 2013–2014 flu season (july 2013–june 2014) from the Google flu trend Web site. Moreover, daily maximum and minimum temperature averaged over weather stations within each continental state is obtained by means of the US historical climatology network. The daily temperature variation (MDTV) is considered as explanatory covariate, being the difference between the daily maximum and daily minimum. The temperature fluctuation is aggregated to the same resolution as the flu activity data by taking the MDTV within each week. Only 42 states are considered due to missed records. We refer to \cite{wang2017unified} for more details.\\

\begin{figure}[htb]
	\centering
	\includegraphics[width=\textwidth]{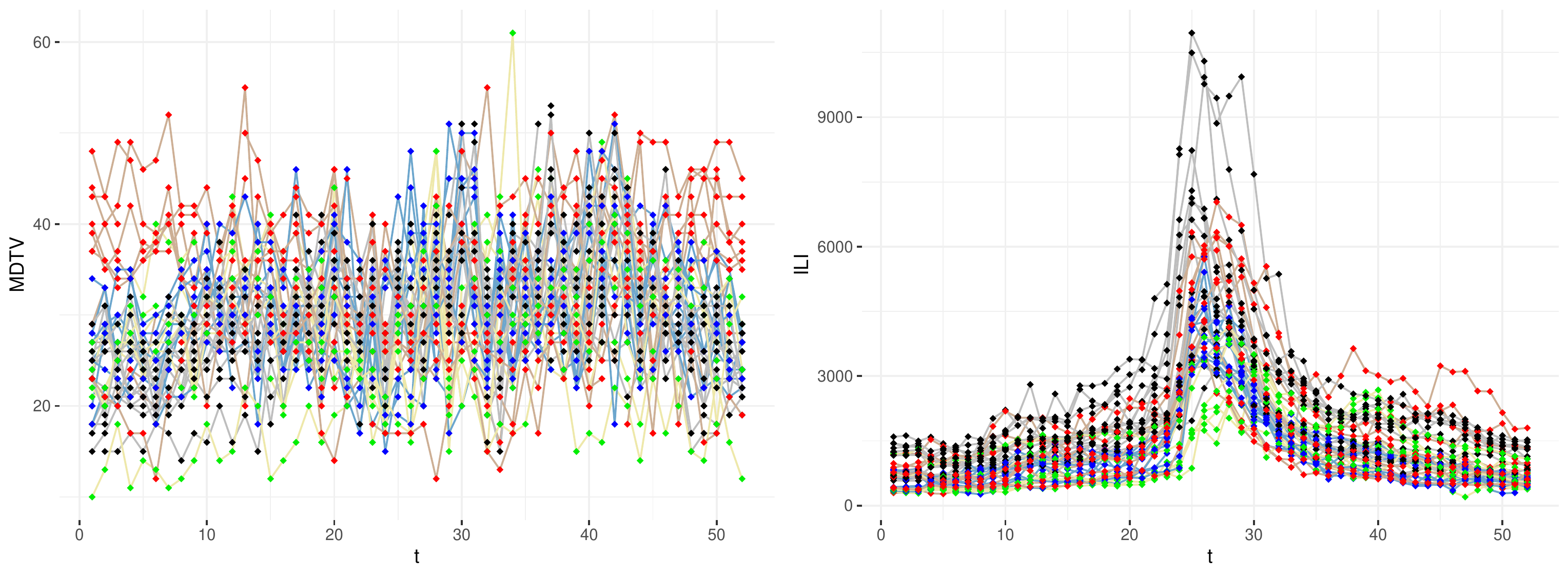}
	\caption{MDTV (left) and flu activity or ILI (right) data in terms of their corresponding regions: north-east (\textcolor{green}{$\bullet$}), midwest (\textcolor{blue}{$\bullet$}), south (\textcolor{black}{$\bullet$}) and west (\textcolor{red}{$\bullet$}).}
	\label{flu_data_plot}
\end{figure}

The original dates from july 1st, 2013, to june 30th, 2014, were numbered by integers from 1 to 365. Then, time $t$ is rescaled to the $[0,1]$ interval by dividing the numbers by 365. Besides, we consider regional effects dividing the data in four sets in terms of midwest, north-east, south or west region to study them separately. Following \cite{wang2017unified}, the ILI percentage and MDTV are standardized at each time point $t$ by dividing the variables by their root mean squares. Data of study is showed in Figure \ref{flu_data_plot} separating this by the considered regions.\\

Therefore, we want to test if the MDTV have relevant information in the flu tendency modeling in terms of the four considered regions. For this aim, we can apply a global test for each region. Results of dependence tests are displayed in Table \ref{p_values_flu_U}. In view that all p-values are greater than $0.1$ we can conclude that we do not have evidences to reject the null hypothesis of mean conditional independence for usual levels as $10\%$. As a result, the MDTV does not play an important role in the ILI modeling no matter the US region. We can argue that maybe the regional effect is not important and we must consider the data jointly. For this purpose, a global test considering all the states is implemented, obtaining a p-value$\approx 0$. This bring outs the fact that we have evidences to reject the conditional mean independence between MDTV and ILI. As a result, MDTV supports useful information to explain the ILI behavior but this is equal in the four considered regions, so a distinction does not make sense.\\


\begin{table}[htb] 
	\centering
	\small 
	\begin{tabular}{c cccc }
		\toprule
		\textbf{p-value} & midwest & north-east & south & west \\
		\hline
		\multicolumn{1}{c}{\rule{0pt}{0.4cm}} & 0.106 & 0.761 & 0.623 & 0.667   \\
		\bottomrule
	\end{tabular}
	\caption{P-values of the MDD-based tests for the different regions.}
	\label{p_values_flu_U}
\end{table}


Our results agree with the ones of \cite{wang2017unified}. First, they reject the location effect for the linear model formulation. Secondly, they obtain that the MDTV covariate can be eluded from the linear model for a $10\%$ significance level but not for the $5\%$ (p-value=0.052). Thus, they have moderately significant evidence that the MDTV plays a role in the ILI explanation at least in the linear context. It is important to remark that differences may be due to the fact that they are assuming linearity in their regression model jointly with the application of a data preprocessing to remove spatial correlations.\\

\subsection{Bike sharing data}

\begin{figure}[htb]
	\centering
	\includegraphics[width=\textwidth]{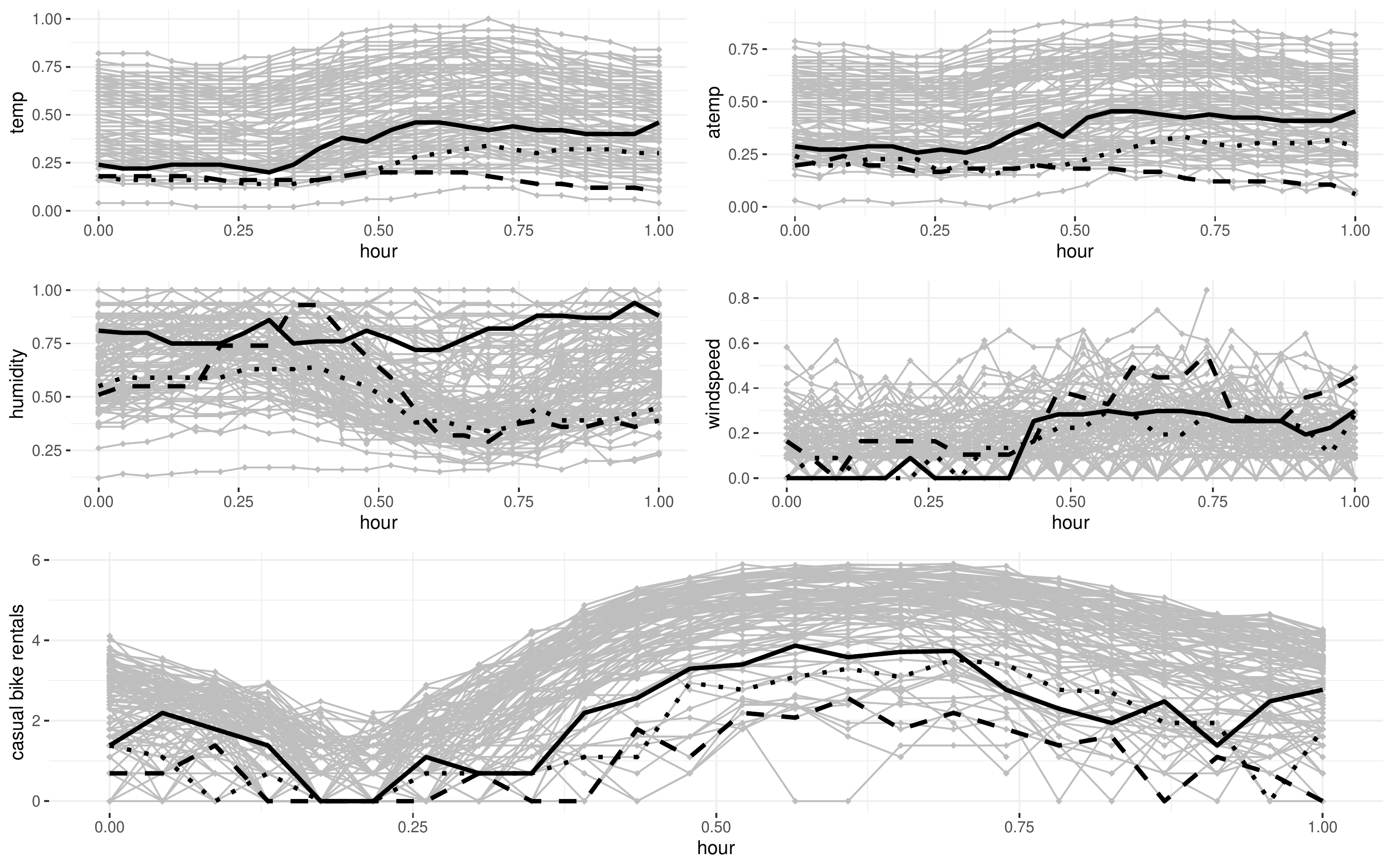}
	\caption{Daily temperature (temp), feeling temperature (atemp), humidity, wind speed and casual bike rentals on an hourly basis in Washington D.C. on Saturdays. }
	\label{bike_sharing}
\end{figure}

Next, a bike sharing dataset of the Washington, D.C., program is analyzed. This is introduced in \cite{Fanaee2014}. The data is obtained daily by the Capital bikeshare system in Washington, D.C., from 1 January 2011 to 31 December
2012. The aim is to explain the number of casual rentals in terms of meteorological covariates. As a result, this dataset contains information on casual bike rentals in the cited period along with other meteorological variables such as temperature in Celsius (temp), feels-like temperature in Celsius (atemp), relative humidity in percentage (humidity) and wind speed in Km/h (windspeed) on an hourly basis. In particular, only the data corresponding with Saturdays are considered because the dynamics changes between working and weekend days, resulting in a total of 105 Saturdays barring some exceptions (8 missing). All covariates are normalized by formula $(t-t_{\min})/(t_{\max}-t_{\min})$ in case of temp and atemp, and dividing by the maximum for humidity and windspeed. In order to correct the skewness of the hourly bike rentals distribution ($Y(t)$) we apply a log transformation considering as response variable $Y(t)=log(Y(t)+1)$. These are showed in Figure \ref{bike_sharing}.\\

First of all, in order to apply our procedure, the missing data is recovered by means of splines interpolation as it was described in Section \ref{subsec:dif_inst}. Then, once we have a total of $n=105$ data points at each time instant, the global significance MDD-based test is performed. We obtain a p-value$=0$, which rejects the null hypothesis of independence for usual significant values as the $5\%$ or the $1\%$.\\

Next, we perform partial tests to detect if any of the four considered covariates (temp, atemp, humidity and windspeed) can be excluded from the model. We obtain p-values of $0$, $0$, $0.007$ and $0.001$ for temperature (temp), feels-like temperature (atemp), relative humidity (humidity) and wind speed (windspeed), respectively. Thus, we can claim that all of them have an impact on the number of casual rentals at significance levels as the $1\%$. This last agrees with other studies as the one of \cite{Ghosal2022}, where different covariates are selected by the considered penalizations. In an overview of their results, each covariate is selected at least two times over the five considered procedures. As a result, all of them seem to play a relevant role separately.\\


\section{Discussion}\label{sec:conclusions}

We propose novel significance tests for the additive functional concurrent model, which collects a wide range of different structures between functional covariates and response. As a result, we test the relevance of a subset of covariates to model the response, including the complete set or partial tests to apply covariates screening. This allow us to detect irrelevant variables and to reduce the problem dimensionality, facilitating the subsequent estimation procedure. For this aim, test statistics based on MDD-ideas are constructed taking under consideration all observed time instants, resulting in global tests to determine the covariates relevance over the complete trajectory. In contrast with existing methodology in literature for significance tests in the concurrent model, as the FLCM (\cite{GHOSAL2022Score}) or the ANFCM (\cite{Kim2018Additive}) procedures among others, our approach has the novel property that there is no need of a preliminary estimation of the model structure. Furthermore, no tuning parameters are involved in contrast with previous methodologies. Instead, it is only needed to compute a $U$-statistic version of the MDD to be able to apply the tests. Using theory of $U$-statistics, good properties of this estimator are guaranteed in practice, as its unbiasedness. Furthermore, its asymptotic distribution is obtained both, under the null and local alternative hypothesis. Eventually, bootstrap procedures are implemented to obtain its p-values in practice.\\

Our new tests have displayed good performance in the linear formulation as in nonlinear structure as can be appreciated by means of the results of scenarios A and B considered in the simulation study, respectively. These procedures are well calibrated under the null hypothesis of no effect, tending to the significance level as the sample size increases. Moreover, they have power under alternatives, which is deduced observing that p-values tend to the unit as sample size increases when associated covariates have an effect on the response. Besides, they seem to perform well in real data sets too. We have exemplified this last by means of three datasets which have been previously studied, comparing our results with existing literature and obtaining similar results when these are comparable. As a result, we have a pretty transversal tool to detect additive effects in the concurrent model framework without the need of previous assumptions or model structure estimation. Furthermore, it is interesting to notice that all these ideas could be translated to the conditional quantile dependence testing in the concurrent model framework. For this purpose, a similar development would be enough following the guidelines and adapting the ideas of Section 3 in \cite{Zhang2018}.\\

In terms of performance comparison with existing literature, we have compared our methodology with \cite{GHOSAL2022Score} (FLCM) and \cite{Kim2018Additive} (ANFCM) algorithms in the linear and additive model framework, respectively. Based on the results, we can claim that our procedure is quite competitive, even when the FLCM and ANFCM have the advantage that they are implemented assuming the correct model structure and an optimal number of basis components is employed. In contrast, simulating under a more functional scenario and avoiding only trigonometric expressions in the model, our procedure clearly outperforms both of them. Besides, other disadvantage for their procedures is that $m(t,X(t))$ is unknown in practice and a misguided assumption of the model structure could lead to wrong results. Moreover, as it is discussed in \cite{GHOSAL2022Score} and \cite{Kim2018Additive}, the proper selection of the number of basis components is quite difficult in practice and it is still an open problem. This plays the role of tuning parameter, so an appropriate value is needed to guarantee a good adjustment. In contrast, our proposal has the novelty that this does not required about previous estimation or tuning parameters selection bridging a gap and solving these problems.\\

One limitation of the present form of our test is that this only admits the study of continuous covariates. This is a common restriction in the concurrent model significance tests framework, as we can appreciate in the works of \cite{GHOSAL2022Score} and \cite{Kim2018Additive}. If we want to be able to include categorical variables, as it is done in other tests as in \cite{wang2017unified}, a different metric is needed to correctly define the $U$-statistic of the MDD test. Some solutions for this problem have been already proposed for the distance covariance approach in the presence of noncontinuous variables, so similar ideas can be translated to the MDD context to solve this issue. We leave this topic for future research.\\

Other drawback is related with the observed time instants disposal. We need to monitor the same number of curves on each time instant to be able to construct our statistic. This translates in $n_t=n$ observed curves points for all $t\in\mathcal{D}_t$. If only a small number of points are missed, we have seen that we can impute them by means of interpolation. However, if we are in a sparse context where each curve is observed in a different number of time points, where they may not coincide with each other, it is quite difficult to preprocess the data to obtain our starting point. In conclusion, a new methodology is needed for these scenarios based on different dependence measures. This is a really interesting area of study for next work. In relation with this, an additional drawback is the statistics computational time, being of the order of $O(n(n-1)(n-2)(n-3)\mathcal{T})$ operations. Then, this procedure is quite competitive for ``moderate'' values of $n$ and $\mathcal{T}$, however, for great values of these quantities, specially for large values $n$, its computational cost is pretty high. As a consequence, simplification techniques in the number of required operations are of interest to make the procedure more tractable.\\

Eventually, we remark that, because of the statistics structures, the tests collect only additive effects. Although this formulation embraces a huge variety of different structures, this does not consider some complex relations like interactions. Nevertheless, we think that by means of projections our ideas can be extended to the general concurrent model formulation, where all possible relations are considered. This is a complete new line for our future research.\\

\section*{Acknowledgment}

The research of Laura Freijeiro-Gonz\'alez is supported by the Conseller\'ia de Cultura, 
Educaci\'on e Ordenaci\'on Universitaria along with the Consellería de Econom\'ia, Emprego 
e Industria of the Xunta de Galicia (project ED481A-2018/264). Laura Freijeiro-Gonz\'alez, 
Wenceslao Gonz\'alez-Manteiga and Manuel Febrero-Bande acknowledged the support from Project PID2020-116587GB-I00 funded by MCIN/AEI/10.13039/501100011033 and by “ERDF A way of making Europe” and the Competitive Reference 
Groups 2021–2024 (ED431C 2021/24) from the Xunta de Galicia through the ERDF. Besides, we acknowledge to the Centro de Supercomputación de Galicia (CESGA) for computational resources.\\


\bibliographystyle{apalike}
\bibliography{Libreria_concurrent}

\pagebreak		
\appendix
\section{Appendix}\label{apend}

In this section we collect the proofs of the main results of the paper jointly with competitors simulation details and extra graphics.\\

We remind that operators $\mathbb{E}[\cdot]$, $\Var[\cdot]$, $\Cov[\cdot, \cdot]$ and $\text{tr}[\cdot]$ apply for the expectation, variance, covariance and matrix trace, respectively. Terms denoted as $\tilde{A}$ represents an integrated version in the $\mathcal{D}_t$ domain, i.e. $\tilde{A}=\int_{\mathcal{D}_t} A(t) dt$. Symbols $\overline{A}$ and $\overline{\overline{A}}$ are the simple and double $\mathcal{U}$-centering versions introduced in Section \ref{sec:MDD} of the main document. When both situations arise together, for example $\widetilde{\overline{\overline{B}}}$, this means the integrated version of the double centered (in this example) term. We keep notation $\dot{U}(X(t),X'(t))$ to say that this term changes now with the sample size $n$ and the dimension of the considered covariates, $d$.

\subsection{Unbiasedness of $\widetilde{MDD}_{n}^2(Y(t)|_{X_j(t)})$ } \label{append:unbiasedness}

First of all, making use of expression (\ref{MDD_UV}), we remind that
\begin{equation*}
	\begin{split}
		\widetilde{MDD}^2(Y(t)|_{X_j(t)}) =& \mathbb{E} \left[ \widetilde{ K\left( X_j(t),X'_j(t) \right) L(Y(t),Y'(t)) }  \right] + \mathbb{E} \left[ \widetilde{ K\left( X_j(t),X'_j(t) \right)} \right] \mathbb{E} \left[ \widetilde{ L(Y(t),Y'(t)) }  \right]\\
		&-2 \mathbb{E} \left[ \widetilde{ K\left( X_j(t),X'_j(t) \right) L(Y(t),Y''(t)) }  \right]
	\end{split}
\end{equation*}
with $\widetilde{ K\left( X_j(t),X'_j(t) \right)}=\int_{\mathcal{D}_t} |X_j(t)-X'_j(t)| dt$ and $\widetilde{ L(Y(t),Y'(t)) }= 1/2\int_{\mathcal{D}_t} \left(Y(t)-Y'(t)\right)^2 dt$.\\

Now, applying $\mathcal{U}$-centering properties, it is verified that
\begin{equation*}
	\begin{split}
		&\sum_{i\neq l} \int_{\mathcal{D}_t} \left( A_{il}(t) \right)_j  \overline{\overline{B}}_{il}(t) dt\\
		&=\sum_{i\neq l} \int_{\mathcal{D}_t} \left( A_{il}(t) \right)_j  \left(B_{il}(t) -\frac{1}{n-2} \sum_{q=1}^{n}  B_{iq}(t)  -\frac{1}{n-2} \sum_{r=1}^{n}  B_{rl}(t)  + \frac{1}{(n-1)(n-2)} \sum_{q,r=1}^{n} B_{qr}(t)  \right)dt\\
		&=\sum_{i\neq l} \int_{\mathcal{D}_t} \left( A_{il}(t) \right)_j B_{il}(t)  dt -\frac{1}{n-2} \sum_{i\neq l} \sum_{q=1}^{n} \int_{\mathcal{D}_t} \left( A_{il}(t) \right)_j  B_{iq}(t)  dt \\
		& \quad -\frac{1}{n-2} \sum_{i\neq l} \sum_{r=1}^{n} \int_{\mathcal{D}_t} \left( A_{il}(t) \right)_j  B_{rl}(t)  dt+ \frac{1}{(n-1)(n-2)} \sum_{i,l=1}^n \sum_{q,r=1}^{n}  \int_{\mathcal{D}_t} \left( A_{il}(t) \right)_j  B_{qr}(t)  dt\\
		&=  \text{tr}\left[ \left( \widetilde{AB} \right)_j \right] +\frac{\mathbf{1}^\top_n \left( \tilde{A} \right)_j \mathbf{1}_n \mathbf{1}^\top_n  \tilde{B}  \mathbf{1}_n}{(n-1)(n-2)}-\frac{2 \mathbf{1}^\top_n \left( \widetilde{AB} \right)_j \mathbf{1}_n}{(n-2)}
	\end{split}
\end{equation*}
where $\mathbf{1}_n\in\mathbb{R}^n$ is a vector of ones and $\tilde{\left( A\right)}_j=\int_{\mathcal{D}_t} \left( A(t) \right)_j dt$.\\

Using Lemma 1 of \cite{ParkShaoYao2015}, $n(n-3)\left( (\overline{A} )_j \cdot \overline{B} \right)= \sum_{i\neq l} (\overline{A}_{il} )_j  \overline{B}_{il}= \sum_{i\neq l} (A_{il} )_j  \widetilde{\overline{\overline{B}}}_{il}$. Then, using the fact that $\tilde{B}_{ii}=0$ it is verified that $\widetilde{\overline{\overline{B}}}_{il}=\widetilde{\overline{B}}_{il}$ and we have
\begin{equation*}
	\widetilde{MDD}_{n}^2(Y(t)|_{X_j(t)})= \frac{1}{n(n-3)} \left( \text{tr}\left[ \left( \widetilde{AB} \right)_j \right] +\frac{\mathbf{1}^\top_n \left( \tilde{A} \right)_j \mathbf{1}_n \mathbf{1}^\top_n  \overline{B}  \mathbf{1}_n}{(n-1)(n-2)}-\frac{2 \mathbf{1}^\top_n \left( \widetilde{AB} \right)_j \mathbf{1}_n}{(n-2)} \right)
\end{equation*}

Denote by $(n)_k=n!/(n-k)!$ and $I_k^n$ the $k$-tuples of indices $\{1,\dots,n\}$ without replacement. Then, it can be seen that 
\begin{equation*}
	\begin{split}
		&(n)_2^{-1} \mathbb{E} \left[ \sum_{(i,l)\in I_2^n} \int_{\mathcal{D}_t} \left( A_{il}(t) \right)_j  B_{il}(t)  dt \right] = (n)_2^{-1} \mathbb{E} \left[ \text{tr}\left[ \left( \widetilde{AB} \right)_j \right] \right] = \mathbb{E} \left[ \widetilde{ K\left( X_j(t),X'_j(t) \right) L(Y(t),Y'(t)) }  \right] \\
		&(n)_4^{-1} \mathbb{E} \left[ \sum_{(i,l,q,r)\in I_4^n} \int_{\mathcal{D}_t} \left( A_{il}(t) \right)_j  B_{qr}(t)  dt \right]= (n)_4^{-1} \mathbb{E}\left[ \mathbf{1}^\top_n \left( \tilde{A} \right)_j \mathbf{1}_n \mathbf{1}^\top_n  \overline{B}  \mathbf{1}_n -4\mathbf{1}^\top_n \left( \widetilde{AB} \right)_j \mathbf{1}_n \right.\\
		&\hspace{7.2cm} \left. +2\text{tr}\left[ \left( \widetilde{AB} \right)_j \right] \right]= \mathbb{E} \left[ \widetilde{ K\left( X_j(t),X'_j(t) \right)} \right] \mathbb{E} \left[ \widetilde{ L(Y(t),Y'(t)) }  \right]\\
		&(n)_3^{-1} \mathbb{E} \left[ \sum_{(i,l,q)\in I_3^n} \int_{\mathcal{D}_t} \left( A_{il}(t) \right)_j  B_{iq}(t)  dt \right]= (n)_3^{-1} \mathbb{E}\left[ \mathbf{1}^\top_n \left( \widetilde{AB} \right)_j \mathbf{1}_n -\text{tr}\left[ \left( \widetilde{AB} \right)_j \right] \right]\\
		&\hspace{7.2cm} = \mathbb{E} \left[ \widetilde{ K\left( X_j(t),X'_j(t) \right) L(Y(t),Y''(t)) }  \right]
 	\end{split}
\end{equation*}

As a consequence, seeing that
\begin{equation*}
	\begin{split}
			\widetilde{MDD}_{n}^2(Y(t)|_{X_j(t)}) =& (n)_2^{-1} \sum_{(i,l)\in I_2^n} \int_{\mathcal{D}_t} \left( A_{il}(t) \right)_j  B_{il}(t)  dt + (n)_4^{-1} \sum_{(i,l,q,r)\in I_4^n} \int_{\mathcal{D}_t} \left( A_{il}(t) \right)_j  B_{qr}(t)  dt\\
			&-2 (n)_3^{-1} \sum_{(i,l,q)\in I_3^n} \int_{\mathcal{D}_t} \left( A_{il}(t) \right)_j  B_{iq}(t)  dt
	\end{split}
\end{equation*}
this is clearly an unbiased estimator of $\widetilde{MDD}^2(Y(t)|_{X_j(t)})$.\\

\subsection{Hoeffding decomposition}\label{append:Hoeffding}

Let $\widetilde{h_c(w_1,\dots,w_c)}=\mathbb{E}\left[ \int_{\mathcal{D}_t} h(w_1,\dots,w_c,Z_{c+1}(t),\dots,Z_4(t)) dt \right]$ for $c=1,2,3,4$ and $Z_i(t)=(X_i(t),Y_i(t))\stackbin{d}{=}(X(t),Y(t))$, where $h(\cdot)$ is defined in (\ref{h_tilde}) and $\stackbin{d}{=}$ means equal in distribution. Let $w=(x,y)$, $w'=(x',y')$, $w''=(x'',y'')$ and $w'''=(x''',y''')$, where $x,x',x'',x'''\in \mathbb{R}^p$ and $y,y',y'',y'''\in\mathbb{R}$. Besides let $Z'(t)=(X'(t),Y'(t))$, $Z''(t)=(X''(t),Y''(t))$ and $Z'''(t)=(X'''(t),Y'''(t))$ independent copies of $Z(t)=(X(t),Y(t))$. We define $\widetilde{U(x,x')}=\int_{\mathcal{D}_t} \mathbb{E}[K(x,X'(t))]dt$ $+\int_{\mathcal{D}_t} \mathbb{E}[K(X(t),x')]dt$ $-\int_{\mathcal{D}_t} K(x,x') dt$ $-\int_{\mathcal{D}_t}\mathbb{E}[K( X(t),X'(t) )]dt$ and $\widetilde{V(y,y')}=\int_{\mathcal{D}_t} (y-\mu_Y)(y'-\mu_Y)dt$ taking $\mu_Y=\mathbb{E}[Y(t)]$. Then, we obtain that
\begin{equation*}
	\widetilde{h_1(w)}=\frac{1}{2} \left\lbrace \mathbb{E} \left[\widetilde{U(x,X(t))} \widetilde{V(y,Y(t))} \right] + \widetilde{MDD}^2\left(Y(t)|_{X(t)}\right) \right\rbrace
\end{equation*}
and
\begin{equation*}
	\begin{split}
		&\widetilde{h_2(w,w')}\\
		&=\frac{1}{6} \left\lbrace  \widetilde{U(x,x')}\widetilde{V(y,y')} + \widetilde{MDD}^2(Y(t)|_{X(t)}) + \mathbb{E}\left[\widetilde{U(x,X(t))}\widetilde{V(y,Y(t))}\right] + \mathbb{E}\left[\widetilde{U(x',X(t))}\widetilde{V(y',Y(t))} \right] \right. \\
		& \quad + \left. \mathbb{E}\left[ \left(\widetilde{U(x,X(t))}-\widetilde{U(x',X(t))}\right)\left(\widetilde{V(y,Y(t))}-\widetilde{V(y',Y(t))}\right) \right]		\right\rbrace
	\end{split}
\end{equation*}

Besides, we have
\begin{equation*}
	\begin{split}
		\widetilde{h_3(w,w',w'')}&=\frac{1}{12} \left\lbrace \left(2\widetilde{U(x,x')}-\widetilde{U(x',x'')}-\widetilde{U(x,x'')}\right)\widetilde{V(y,y')} \right.\\
		 & \quad+\left(2\widetilde{U(x,x'')}-\widetilde{U(x,x')}-\widetilde{U(x',x'')}\right) \widetilde{V(y,y'')} \\
		 & \quad+ \left(2\widetilde{U(x',x'')}-\widetilde{U(x,x')}-\widetilde{U(x,x'')}\right) \widetilde{V(y',y'')} \\
		 & \quad+\mathbb{E}\left[\left(2\widetilde{U(x,X(t))}-\widetilde{U(x',X(t))}-\widetilde{U(x'',X(t))}\right) \widetilde{V(y,Y(t))}\right] \\
		 & \quad+\mathbb{E}\left[\left(2\widetilde{U(x',X(t))}-\widetilde{U(x,X(t))}-\widetilde{U(x'',X(t))}\right) \widetilde{V(y',Y(t))}\right] \\
		 & \quad+ \left. \mathbb{E}\left[ \left(2\widetilde{U(x'',X(t))}-\widetilde{U(x,X(t))}-\widetilde{U(x',X(t))}\right) \widetilde{V(y'',Y(t))}\right] \right\rbrace
	\end{split}
\end{equation*}
and
\begin{equation*}
	\begin{split}
		&\widetilde{h_4(w,w',w'',w''')} \\
		&=\frac{1}{12} \left\lbrace  \left(2\widetilde{U(x,x')}+2\widetilde{U(x'',x''')}-\widetilde{U(x,x'')}-\widetilde{U(x,x''')}-\widetilde{U(x',x'')}-\widetilde{U(x',x''')} \right) \left(\widetilde{V(y,y')}+\widetilde{V(y'',y''')} \right) \right. \\
		&\quad + \left(2\widetilde{U(x,x'')}+2\widetilde{U(x',x''')}-\widetilde{U(x,x')}-\widetilde{U(x,x''')}-\widetilde{U(x'',x')}-\widetilde{U(x'',x''')}\right) \left(\widetilde{V(y,y'')}+\widetilde{V(y',y''')} \right) \\
		&\quad + \left. \left(2\widetilde{U(x,x''')}+2\widetilde{U(x'',x')}-\widetilde{U(x,x'')}-\widetilde{U(x,x')}-\widetilde{U(x''',x'')}-\widetilde{U(x''',x')} \right)\left(\widetilde{V(y,y''')}+\widetilde{V(y',y'')}\right)	\right\rbrace
	\end{split}
\end{equation*}

\subsubsection{Analysis under the null hypothesis}
Under the null hypothesis we have that $\mathbb{E} \left[ Y(t) |_{X_j(t)} \right]=\mathbb{E} \left[ Y(t) \right]$ almost surely $\forall  t\in \mathcal{D}_t$ and every $j=1,\dots,p$, which also translates in $\widetilde{MDD}^2(Y(t)|_{X(t)}) =0$. Then, it is verified that $\widetilde{h_1(w)}=0$, $\widetilde{h_2(w,w')}= \widetilde{U(x,x')}\widetilde{V(y,y')}/6$ and
\begin{equation*}
	\begin{split}
		\widetilde{h_3(w,w',w'')}=& \frac{1}{12} \left\lbrace \left( 2\widetilde{U(x,x')}-\widetilde{U(x',x'')}-\widetilde{U(x,x'')} \right) \widetilde{V(y,y')}\right. \\
		& +\left(2\widetilde{U(x,x'')}-\widetilde{U(x,x')}-\widetilde{U(x',x'')} \right) \widetilde{V(y,y'')} \\
		& +\left. \left(2\widetilde{U(x',x'')}-\widetilde{U(x,x')}-\widetilde{U(x,x'')} \right) \widetilde{V(y',y'')}  \right\rbrace
	\end{split}
\end{equation*}

Furthermore, under the null, we can verify that
\begin{equation*}
	\Var[\widetilde{h_2(Z(t),Z'(t))}]=\frac{1}{36} \mathbb{E}\left[\widetilde{U(X(t),X'(t))}^2\widetilde{V(Y(t),Y'(t))}^2 \right] = \frac{1}{36}\tilde{\xi}^2
\end{equation*}
and
\begin{equation*}
	\begin{split}
		\Var&[\widetilde{h_3(Z(t),Z'(t),Z''(t))}]\\
		&=\frac{3}{144} \text{Var}\left[ \left(2\widetilde{U(X(t),X'(t))}-\widetilde{U(X'(t),X''(t))}-\widetilde{U(X(t),X''(t))}\right)\widetilde{V(Y(t),Y'(t))} \right]\\
		&=\frac{3}{144} \left\lbrace 4\tilde{\xi}^2 +2\mathbb{E}\left[\widetilde{U(X(t),X''(t))}^2 \widetilde{V(Y(t),Y'(t))}^2\right] \right.\\
		&\quad \left. +2\mathbb{E}\left[\widetilde{U(X(t),X''(t))}\widetilde{U(X'(t),X''(t))}\widetilde{V(Y(t),Y'(t))}^2\right] \right\rbrace .
	\end{split}
\end{equation*}
Moreover,
\begin{equation*}
	\begin{split}
		\Var&[\widetilde{h_4(Z(t),Z'(t),Z''(t),Z'''(t))}]\\
		&=\frac{6}{144}\mathbb{E} \left[ \widetilde{V(Y(t),Y'(t))}^2 \left( \widetilde{U(X(t),X''(t))}+\widetilde{U(X'(t),X'''(t))} +\widetilde{U(X'(t),X''(t))} \right.  \right.\\
		&\quad + \left. \left. \widetilde{U(X(t),X'''(t))}-2\widetilde{U(X(t),X'(t))} -2\widetilde{U(X''(t),X'''(t))}\right)^2 \right] \\
		&=\frac{1}{6} \left\lbrace \mathbb{E}\left[ \widetilde{V(Y(t),Y'(t))}^2 \widetilde{U(X(t),X''(t))}\widetilde{U(X'(t),X''(t))} \right] + \tilde{\xi}^2 \right.\\
		&\quad + \left. \mathbb{E} \left[ \widetilde{V(Y(t),Y'(t))}^2\widetilde{U(X(t),X''(t))}^2 \right] + \mathbb{E}\left[ \widetilde{V(Y(t),Y'(t))}^2 \right] \mathbb{E}\left[ \widetilde{U(X(t),X''(t))}^2 \right] \right\rbrace
	\end{split}
\end{equation*}\\
	
Using the Cauchy-Schwarz inequality we can obtain that
\begin{equation*}
	\begin{split}
		& \mathbb{E} \left[  \widetilde{U(X(t),X''(t))}\widetilde{U(X'(t),X''(t))}\widetilde{V(Y(t),Y'(t))}^2 \right]\\
		& \leq \left\lbrace \mathbb{E}\left[ \widetilde{U(X(t),X''(t))}^2\widetilde{V(Y(t),Y'(t))}^2 \right] \right\rbrace^{1/2} \left\lbrace \mathbb{E}\left[ \widetilde{U(X'(t),X''(t))}^2\widetilde{V(Y(t),Y'(t))}^2 \right] \right\rbrace^{1/2}\\
		&=\mathbb{E} \left[ \widetilde{U(X(t),X''(t))}^2\widetilde{V(Y(t),Y'(t))}^2 \right].
	\end{split}
\end{equation*}	
	
Besides, under the assumption that
\begin{equation*}
	\begin{split}
		& \frac{\mathbb{E}\left[ \widetilde{U(X(t),X''(t))}^2 \widetilde{V(Y(t),Y'(t))}^2 \right]}{\tilde{\xi}^2}=o(n),\\
		& \frac{\mathbb{E}\left[ \widetilde{V(Y(t),Y'(t))}^2 \right] \mathbb{E}\left[ \widetilde{U(X(t),X'(t))}^2 \right]}{\tilde{\xi}^2}=o(n^2),
	\end{split}
\end{equation*}	
we have
\begin{equation*}
	\widetilde{MDD}_n^2(Y(t)|_{X(t)})=\frac{1}{\binom{n}{2}} \sum_{1\leq i < l \leq n} \widetilde{U(X_{i}(t),X_l(t))} \widetilde{V(Y_i(t),Y_l(t))}+\mathcal{R}_n,
\end{equation*}
where $\mathcal{R}_n$ is the remainder term which is asymptotically negligible (see \cite{Serfling1980}).

\subsubsection{Analysis under local alternatives}

We consider the case where $\widetilde{MDD}^2(Y(t)|_{X(t)})$ is nonzero, i.e.,
the conditional mean of $Y(t)$ may depend on $X(t)$. Recall that $\widetilde{L(x, y)} = \mathbb{E}\left[\widetilde{U(x,X(t))} \widetilde{V(y,Y(t))}\right]$. Under the assumption that
\begin{equation}\label{assum_var1_zero}
	\Var\left[ \widetilde{L(X(t), Y(t))} \right]=o(n^{-1}\tilde{\xi}^2), \quad \Var\left[ \widetilde{L(X(t), Y'(t))} \right]=o(\tilde{\xi}^2),
\end{equation}
we get
\begin{equation*}
	\Var\left[ \widetilde{h_1} \right]=o(n^{-1}\tilde{\xi}^2), \quad \Var\left[ \widetilde{h_2} \right]=\frac{\tilde{\xi}^2}{36}(1+o(1))
\end{equation*}
which means that $\Var\left[ \widetilde{h_1} \right]$ tends to zero as $n$ increases and $\Var\left[ \widetilde{h_2} \right]$ is always positive and nonnull.\\

Moreover,
\begin{equation*}
	\Var\left[ \widetilde{h_3(Z(t),Z'(t),Z''(t))} \right] \leq C \left\lbrace \tilde{\xi}^2 + \mathbb{E}\left[ \widetilde{U(X(t),X''(t))}^2 \widetilde{V(Y(t),Y'(t))}^2 \right] \right\rbrace
\end{equation*}
and
\begin{equation*}
	\begin{split}
		\Var&\left[ \widetilde{h_4(Z(t),Z'(t),Z''(t),Z'''(t))} \right]\\
	 	&\leq C' \left\lbrace \mathbb{E}\left[ \widetilde{V(Y(t),Y'(t))}^2 \widetilde{U(X(t),X''(t))}^2 \right] + \mathbb{E}\left[ \widetilde{V(Y(t),Y'(t))}^2 \right] \mathbb{E}\left[ \widetilde{U(X(t),X'(t))}^2 \right] + \tilde{\xi}^2  \right\rbrace
	 	\end{split}
\end{equation*}
for some constants $C, C'\geq 0$.\\

Then, under assumption (\ref{assum_var1_zero}),
\begin{equation*}
	\widetilde{MDD}^2_n(Y(t)|_{X(t)})-\widetilde{MDD}^2(Y(t)|_{X(t)})=\frac{1}{\binom{n}{2}} \sum_{1\leq i <l\leq n} \widetilde{U(X_{i}(t),X_l(t))} \widetilde{V(Y_i(t),Y_l(t))}+\mathcal{R}_n.
\end{equation*}

Applying the above arguments to $\sum_{j\in D} \widetilde{MDD}^2_n(Y(t)|_{X_j(t)})$, it can be seen that
\begin{equation*}
	\sum_{j\in D} \left\lbrace \widetilde{MDD}^2_n(Y(t)|_{X_j(t)})-\widetilde{MDD}^2(Y(t)|_{X_j(t)}) \right\rbrace =\frac{1}{\binom{n}{2}} \sum_{1\leq i <l\leq n} \widetilde{\dot{U}(X_{i}(t),X_l(t))} \widetilde{V(Y_i(t),Y_l(t))}+ \sum_{j\in D} (\mathcal{R}_n)_j.
\end{equation*}
where the kernel $\widetilde{\dot{U}(X(t),X'(t))}=\sum_{j\in D} \widetilde{U_j(x_j(t),x'_j(t))}$ is changing now with $(n,d)$. Next, to provide that the remainder term $\sum_{j\in D} (\mathcal{R}_n)_j$ is asymptotically
negligible we assume that 
\begin{equation*}
	\begin{split}
		&\frac{\mathbb{E}\left[ \widetilde{\dot{U}(X(t),X''(t))}^2 \widetilde{V(Y(t),Y'(t))}^2 \right]}{\tilde{S}^2}=o(n),\\
		&\frac{\mathbb{E}\left[ \widetilde{\dot{U}(X(t),X'(t))}^2 \right] \mathbb{E}\left[ \widetilde{V(Y(t),Y'(t))}^2 \right]}{\tilde{S}^2}=o(n^2),\\
		& \text{Var}\left[ \widetilde{\dot{L}(X(t),Y(t))} \right]=o(n^{-1}\tilde{S}^2), \quad \text{Var}\left[ \widetilde{\dot{L}(X(t),Y'(t))} \right]=o(\tilde{S}^2)
	\end{split}
\end{equation*}
with $ \widetilde{\dot{L}(x,y)}=\mathbb{E}\left[\widetilde{\dot{U}(x,X(t))} \widetilde{V(y,Y(t))}\right]$.\\

\subsection{Asymptotic normality under the null and alternatives}\label{append:normality}

Following \cite{martingale2014} guidelines, we prove the asymptotic normality of $T_D$ making use of the Central Limit Theorem for martingale difference sequences.

\subsubsection{Analysis under $H_0$} 

Define
\begin{equation*}
S_r \coloneqq \sum_{l=2}^{r} \sum_{i=1}^{r-1} \widetilde{\dot{U}(X_i(t),X_l(t))} \widetilde{V(Y_i(t),Y_l(t))}= \sum_{l=2}^{r} \sum_{i=1}^{r-1} \widetilde{H(Z_i(t),Z_l(t))}
\end{equation*}

and the filtration $\mathcal{F}_r=\sigma\{Z_1(t),Z_2(t),\dots,Z_r(t)\}$ with $Z_i(t)=(X_i(t),Y_i(t))$. Then, $S_r$ is adaptive to $\mathcal{F}_r$ and is a mean-zero martingale sequence verifying $\mathbb{E}[S_r]=0$ and
\begin{equation*}
	\mathbb{E}[S_{r'}|_{\mathcal{F}_r}]= S_r + \sum_{l=r+1}^{r'} \sum_{i=1}^{l-1} \mathbb{E}\left[ \mathbb{E}\left[ \widetilde{\dot{U}(X_i(t),X_l(t))} \widetilde{V(Y_i(t),Y_l(t))} |_{\mathcal{F}_r, X_i(t), X_l(t)} \right] |_{\mathcal{F}_r} \right] = S_r
\end{equation*}
for $r'\geq r$. Thus, by Corollary 3.1 of \cite{hall1980martingale} we can guarantee the asymptotic normality of $T_D$ if the conditions (\ref{norm_cond_1}) and (\ref{norm_cond_2}) are verified. Specifically, defining $\mathcal{W}_l=\sum_{i=1}^{l-1} \widetilde{H(Z_i(t),Z_l(t))}$, it is sufficient to see that 
\begin{equation}\label{norm_cond_1}
	\sum_{l=1}^{n} B^{-2} \mathbb{E}\left[ \mathcal{W}^2_l \mathbb{I}(|\mathcal{W}_l|>\varepsilon B)|_{\mathcal{F}_{l-1}} \right] \stackbin{p}{\longrightarrow} 0
\end{equation}
for $B$ such that 
\begin{equation}\label{norm_cond_2}
	\sum_{l=1}^{n} \mathbb{E}\left[ \mathcal{W}^2_l|_{\mathcal{F}_{l-1}} \right]/B^2\stackbin{p}{\longrightarrow} C >0.
\end{equation}

We start proving that (\ref{norm_cond_2}) is verified taking $B^2=n(n-1)\tilde{S}^2/2$ and $C=1$. This translates in proving that 
\begin{equation}\label{norm_cond_2_bis}
	\frac{2}{n(n-1)\tilde{S}^2}\sum_{l=1}^{n} \mathbb{E}\left[ \mathcal{W}^2_l|_{\mathcal{F}_{l-1}} \right]\stackbin{p}{\longrightarrow} 1.
\end{equation}

For this purpose, note that
\begin{equation*}
	\mathbb{E}[\mathcal{W}_l^2|_{\mathcal{F}_{l-1}}]=\mathbb{E} \left[ \sum_{i,k=1}^{l-1} \widetilde{H(Z_i(t),Z_l(t))} \widetilde{H(Z_k(t),Z_l(t))} |_{\mathcal{F}_{l-1}} \right] = \sum_{i,k=1}^{l-1} \widetilde{G(Z_i(t),Z_k(t))},
\end{equation*}
and
\begin{equation}\label{S_expre}
	\begin{split}
		&\frac{2}{n(n-1)}\sum_{l=2}^{n} \mathbb{E}[\mathcal{W}_l^2]\\ &=\frac{2}{n(n-1)}\sum_{l=2}^{n} \mathbb{E}\left[ \sum_{i,k=1}^{l-1} \int_{\mathcal{D}_t} (Y_i(t)-\mu(t))(Y_k(t)-\mu(t))(Y_l(t)-\mu(t))^2 \dot{U}(X_i(t),X_l(t)) \dot{U}(X_k(t),X_l(t))dt \right]\\
		&=\frac{2}{n(n-1)}\sum_{l=2}^{n} \mathbb{E}\left[ \sum_{i,k=1}^{l-1} \int_{\mathcal{D}_t} (Y_i(t)-\mu(t))^2(Y_l(t)-\mu(t))^2 \dot{U}(X_i(t),X_l(t))^2 dt\right]\\
		&=\mathbb{E}\left[ \widetilde{V(Y(t),Y'(t))}^2 \widetilde{\dot{U}(X(t),X'(t))}^2 \right] = \mathbb{E} \left[ \widetilde{H(Z(t),Z'(t))}^2 \right]= \tilde{S}^2.
	\end{split}
\end{equation}

Then, defining 
\begin{equation*}
	\begin{split}
		&\mathcal{D}_1= \mathbb{E}\left[ \widetilde{H(Z(t),Z''(t))}^2 \widetilde{H(Z'(t),Z''(t))}^2 \right] - \left( \mathbb{E}\left[ \widetilde{H(Z(t),Z'(t))}^2\right] \right)^2=\Var\left[ \widetilde{G(Z(t),Z(t))} \right]\\
		&\mathcal{D}_2= \mathbb{E} \left[ \widetilde{H(Z(t),Z'(t)} \widetilde{H(Z'(t),Z''(t)} \widetilde{H(Z''(t),Z'''(t)} \widetilde{H(Z'''(t),Z(t)} \right] = \mathbb{E} \left[ \widetilde{G(Z(t),Z'(t))}^2 \right]
	\end{split}
\end{equation*}

we have for $l \geq l'$ that 
\begin{equation*}
	\begin{split}
	\Cov\left[ \mathbb{E}\left[ \mathcal{W}^2_l|_{\mathcal{F}_{l-1}} \right], \mathbb{E}\left[ \mathcal{W}^2_{l'}|_{\mathcal{F}_{l'-1}} \right] \right]&=\sum_{i,k=1}^{l-1} \sum_{i',k'=1}^{l'-1} \Cov\left[ \widetilde{G(Z_i(t),Z_k(t))}, \widetilde{G(Z_{i'}(t),Z_{k'}(t))} \right] \\
	&=(l'-1)\mathcal{D}_1+2(l'-1)(l'-2)\mathcal{D}_2.	
	\end{split}
\end{equation*}

As a result, under the assumption that 
\begin{equation*}
	\frac{ \mathbb{E}\left[ \widetilde{G(Z(t),Z'(t))}^2 \right]}{ \left\lbrace \mathbb{E} \left[ \widetilde{H(Z(t),Z'(t))}^2 \right]  \right\rbrace ^2 } \longrightarrow 0, \quad \quad \frac{ \mathbb{E} \left[ \widetilde{H(Z(t),Z''(t))}^2 \widetilde{H(Z'(t),Z''(t))}^2 \right] }{ n\left\lbrace \mathbb{E} \left[ \widetilde{H(Z(t),Z'(t))}^2 \right]  \right\rbrace ^2 } \longrightarrow 0
\end{equation*}
we have
\begin{equation}\label{o_S4}
	\frac{4}{n^2(n-1)^2}\sum_{l,l'=2}^{n} \Cov\left[ \mathbb{E}[\mathcal{W}_l^2|_{\mathcal{F}_{l-1}}], \mathbb{E}[\mathcal{W}_{l'}^2|_{\mathcal{F}_{l'-1}}]  \right] = O(\mathcal{D}_1/n+\mathcal{D}_2)=o(S^4).
\end{equation}

Next, we make use of result (\ref{o_S4}) to prove (\ref{norm_cond_2_bis}). We notice that convergence in r-mean for $r\geq1$ implies convergence in probability. Then, proving
\begin{equation*}
	\lim_{n\rightarrow \infty}\mathbb{E} \left[ \left( \frac{2}{n(n-1)\tilde{S}^2}\sum_{l=2}^{n}\mathbb{E}[\mathcal{W}_l^2|_{\mathcal{F}_{l-1}}] -1 \right)^2  \right]=0.
\end{equation*}
implies that condition (\ref{norm_cond_2_bis}) is verified.\\

Given that 
\begin{equation*}
	\begin{split}
		&\mathbb{E} \left[ \left( \frac{2}{n(n-1)\tilde{S}^2}\sum_{l=2}^{n}\mathbb{E}[\mathcal{W}_l^2|_{\mathcal{F}_{l-1}}] -1 \right)^2  \right]\\
		&= \mathbb{E} \left[  \frac{4}{n^2(n-1)^2\tilde{S}^4} \left(\sum_{l=2}^{n}\mathbb{E}[\mathcal{W}_l^2|_{\mathcal{F}_{l-1}}]\right)^2 +1 -\frac{4}{n(n-1)\tilde{S}^2} \sum_{l=2}^{n}\mathbb{E}[\mathcal{W}_l^2|_{\mathcal{F}_{l-1}}]  \right]\\
		&\stackbin{(a)}{=}\frac{4}{n^2(n-1)^2\tilde{S}^4} \mathbb{E} \left[ \left(\sum_{l=2}^{n}\mathbb{E}[\mathcal{W}_l^2|_{\mathcal{F}_{l-1}}]\right)^2 \right] +1 -2\\
		&=\frac{4}{n^2(n-1)^2\tilde{S}^4} \mathbb{E} \left[ \left(\sum_{l=2}^{n}\mathbb{E}[\mathcal{W}_l^2|_{\mathcal{F}_{l-1}}]\right)^2 \right] -1
	\end{split}
\end{equation*}
using in $(a)$ that $\mathbb{E}\left[\mathbb{E}[\mathcal{W}_l^2|_{\mathcal{F}_{l-1}}] \right] = \mathbb{E}[\mathcal{W}_l^2]$ and (\ref{S_expre}). Thus, it is enough to see that 
\begin{equation}\label{norm_cond_2_r_mean}
	\frac{4}{n^2(n-1)^2\tilde{S}^4} \mathbb{E} \left[ \left(\sum_{l=2}^{n}\mathbb{E}[\mathcal{W}_l^2|_{\mathcal{F}_{l-1}}]\right)^2 \right] -1 \longrightarrow 0.
\end{equation}

Seeing that 
\begin{equation*}
	\begin{split}
		\frac{4}{n^2(n-1)^2\tilde{S}^4}& \mathbb{E}\left[ \left( \sum_{l=2}^{n} \mathbb{E} \left[ \mathcal{W}_l^2|_{\mathcal{F}_{l-1}} \right] \right)^2 \right]\\
		&= \frac{4}{n^2(n-1)^2\tilde{S}^4} \mathbb{E}\left[ \sum_{l=2}^{n} \left( \mathbb{E} \left[ \mathcal{W}_l^2|_{\mathcal{F}_{l-1}} \right] \right)^2 +2 \sum_{l=2}^{n-1}  \mathbb{E} \left[ \mathcal{W}_l^2|_{\mathcal{F}_{l-1}} \right] \sum_{k=l+1}^{n} \mathbb{E} \left[ \mathcal{W}_k^2|_{\mathcal{F}_{k-1}} \right] \right]\\
		&= \frac{4}{n^2(n-1)^2\tilde{S}^4} \mathbb{E}\left[ \sum_{l,k=2}^{n}  \mathbb{E} \left[ \mathcal{W}_l^2|_{\mathcal{F}_{l-1}} \right] \mathbb{E} \left[ \mathcal{W}_k^2|_{\mathcal{F}_{k-1}} \right] \right]\\
		&= \frac{4}{n^2(n-1)^2\tilde{S}^4}  \sum_{l,k=2}^{n}  \mathbb{E}\left[ \mathbb{E} \left[ \mathcal{W}_l^2|_{\mathcal{F}_{l-1}} \right] \mathbb{E} \left[ \mathcal{W}_k^2|_{\mathcal{F}_{k-1}} \right] \right]
	\end{split}
\end{equation*}
and 
\begin{equation*}
	\begin{split}
			&\frac{4}{n^2(n-1)^2}\sum_{l,k=2}^{n} \Cov\left[ \mathbb{E} \left[ \mathcal{W}_l^2|_{\mathcal{F}_{l-1}} \right], \mathbb{E} \left[ \mathcal{W}_k^2|_{\mathcal{F}_{k-1}} \right] \right]\\
			&=\frac{4}{n^2(n-1)^2}\sum_{l,k=2}^{n} \mathbb{E} \left[ \mathbb{E} \left[ \mathcal{W}_l^2|_{\mathcal{F}_{l-1}} \right] \cdot \mathbb{E} \left[ \mathcal{W}_k^2|_{\mathcal{F}_{k-1}} \right] \right] - \frac{4}{n^2(n-1)^2}\sum_{l,k=2}^{n} \mathbb{E}\left[ \mathbb{E} \left[ \mathcal{W}_l^2|_{\mathcal{F}_{l-1}} \right] \right] \mathbb{E}\left[ \mathbb{E} \left[ \mathcal{W}_k^2|_{\mathcal{F}_{k-1}} \right] \right]  \\			
			&= \frac{4}{n^2(n-1)^2}\sum_{l,k=2}^{n}  \mathbb{E} \left[ \mathbb{E} \left[ \mathcal{W}_l^2|_{\mathcal{F}_{l-1}} \right] \cdot \mathbb{E} \left[ \mathcal{W}_k^2|_{\mathcal{F}_{k-1}} \right] \right] - \frac{4}{n^2(n-1)^2}\sum_{l=2}^{n} \left( \mathbb{E}[\mathcal{W}_l^2] \right)^2\\
			& \stackbin{(a)}{=} \frac{4}{n^2(n-1)^2}\sum_{l,k=2}^{n}  \mathbb{E} \left[ \mathbb{E} \left[ \mathcal{W}_l^2|_{\mathcal{F}_{l-1}} \right] \cdot \mathbb{E} \left[ \mathcal{W}_k^2|_{\mathcal{F}_{k-1}} \right] \right] - \tilde{S}^4\\
			& \stackbin{(b)}{=}o(\tilde{S}^4)
	\end{split}
\end{equation*}
where we apply (\ref{S_expre}) in $(a)$ and (\ref{o_S4}) in $(b)$, we have guaranteed that
\begin{equation*}
	\frac{4}{n^2(n-1)^2\tilde{S}^4}\sum_{l,k=2}^{n}  \mathbb{E} \left[ \mathbb{E} \left[ \mathcal{W}_l^2|_{\mathcal{F}_{l-1}} \right] \cdot \mathbb{E} \left[ \mathcal{W}_k^2|_{\mathcal{F}_{k-1}} \right] \right] - \frac{\tilde{S}^4}{\tilde{S}^4}\longrightarrow 0,
\end{equation*}
which implies (\ref{norm_cond_2_r_mean}). Then, the convergence in $r$-mean is verified taking $r=2$. This proves condition (\ref{norm_cond_2_bis}) which ensures condition (\ref{norm_cond_2}) for $B^2=n(n-1)\tilde{S}^2/2$ and $C=1$.\\

Next, we prove the remainder condition (\ref{norm_cond_1}) for this value of $B$. For this aim, we notice that 
\begin{equation*}
	0\leq \sum_{l=1}^{n} B^{-2} \mathbb{E}\left[ \mathcal{W}^2_l \mathbb{I}(|\mathcal{W}_l|>\varepsilon B)|_{\mathcal{F}_{l-1}} \right] \leq B^{-2-s} \varepsilon^{-s} \sum_{l=1}^{n} \mathbb{E}\left[ |\mathcal{W}_l|^{2+s} |_{\mathcal{F}_{l-1}} \right] \leq B^{-2-s} \sum_{l=1}^{n} \mathbb{E}\left[ |\mathcal{W}_l|^{2+s} |_{\mathcal{F}_{l-1}} \right]
\end{equation*}  
for some $s>0$. We prove that taking $s=2$
\begin{equation*}
	B^{-4} \sum_{l=1}^{n} \mathbb{E}\left[ |\mathcal{W}_l|^{4} |_{\mathcal{F}_{l-1}} \right]\stackbin{p}{\longrightarrow} 0
\end{equation*}
with $B^2=n(n-1)\tilde{S}^2/2$. It this case it is sufficient to show that
\begin{equation*}
	B^{-4} \sum_{l=1}^{n} \mathbb{E}\left[ |\mathcal{W}_l|^{4} \right] \longrightarrow 0.
\end{equation*}

Under the assumption that
\begin{equation*}
	\frac{ \mathbb{E}\left[ \widetilde{H(Z(t),Z'(t))}^4 \right]/n + \mathbb{E} \left[ \widetilde{H(Z(t),Z''(t))}^2 \widetilde{H(Z'(t),Z''(t))}^2 \right] }{ n\left\lbrace \mathbb{E} \left[ \widetilde{H(Z(t),Z'(t))}^2 \right]  \right\rbrace ^2 } \longrightarrow 0,
\end{equation*}
we have 
\begin{equation*}
	\begin{split}
		\sum_{l=2}^{n} \mathbb{E}\left[ |\mathcal{W}_l|^{4} \right]&= \sum_{l=2}^{n} \sum_{i_1, i_2, i_3, i_4=1}^{l-1} \mathbb{E}\left[ \widetilde{H(Z_{i_1}(t),Z_l(t))} \widetilde{H(Z_{i_2}(t),Z_l(t))} \widetilde{H(Z_{i_3}(t),Z_l(t))} \widetilde{H(Z_{i_4}(t),Z_l(t))} \right]\\
		&\stackbin{(a)}{=}\frac{n(n-1)}{2} \mathbb{E}\left[ \widetilde{H(Z(t),Z'(t))}^4 \right] + 3 \sum_{l=2}^{n} \sum_{i_1\not = i_2} \mathbb{E}\left[ \widetilde{H(Z_{i_1}(t),Z_l(t))}^2 \widetilde{H(Z_{i_2}(t),Z_l(t))}^2 \right]\\
		&=\frac{n(n-1)}{2} \mathbb{E}\left[ \widetilde{H(Z(t),Z'(t))}^4 \right] + 3 \sum_{l=2}^{n} (l-1)(l-2) \mathbb{E} \left[ \widetilde{H(Z(t),Z''(t))}^2 \widetilde{H(Z'(t),Z''(t))}^2 \right]\\
		&=o(B^4)
	\end{split}
\end{equation*}
where in $(a)$ we apply $\mathbb{E}\left[ \widetilde{H(Z(t),Z'(t))} \right]=0$. As a result, this last implies the condition (\ref{norm_cond_1}).

\subsubsection{Analysis under local alternatives}

Now, we extend the asymptotic normality results to local alternatives. We refer to Section 1.7 of the supplementary material of \cite{Zhang2018} for some discussion about the local alternative model. Its arguments can be easily extended to our context considering the corresponding integrated versions.\\

Thus, under the assumption that $\Var\left[ \widetilde{\dot{L}(X(t),Y(t))} \right]=o(n^{-1}\tilde{S}^2)$, we have
\begin{equation*}
	\begin{split}
		&\frac{1}{\sqrt{\binom{n}{2}}\tilde{S}} \sum_{1\leq i < l \leq n} \left\lbrace \widetilde{\dot{U}(X_i(t),X_l(t))} \widetilde{V(Y_i(t),Y_l(t))} - \mathbb{E}\left[ \widetilde{\dot{U}(X_i(t),X_l(t))} \widetilde{V(Y_i(t),Y_l(t))} \right] \right\rbrace\\
		&=\frac{1}{\sqrt{\binom{n}{2}}\tilde{S}} \sum_{1\leq i < l \leq n} \widetilde{H^*(Z_i(t),Z_l(t))}\\
		& \quad + \frac{1}{\sqrt{\binom{n}{2}}\tilde{S}} \sum_{1\leq i < l \leq n} \left\lbrace  \widetilde{\dot{L}(X_i(t),Y_i(t))} + \widetilde{\dot{L}(X_l(t),Y_l(t))} -2 \mathbb{E}\left[ \widetilde{\dot{U}(X_i(t),X_l(t))}\widetilde{V(Y_i(t),Y_l(t))} \right]  \right\rbrace \\
		&= \frac{1}{\sqrt{\binom{n}{2}}\tilde{S}} \sum_{1\leq i < l \leq n}   \widetilde{H^*(Z_i(t),Z_l(t))} + o_p(1).
	\end{split}
\end{equation*}

Using similar arguments by replacing $H$ with $H^{*}$, we can show that 
\begin{equation*}
	 \frac{1}{\sqrt{\binom{n}{2}}\tilde{S}} \sum_{1\leq i < l \leq n}   \widetilde{H^*(Z_i(t),Z_l(t))} \stackbin{d}{\longrightarrow} N(0,1),
\end{equation*}
which implies that, using Slutsky theorem,
\begin{equation*}
	\frac{1}{\sqrt{\binom{n}{2}}\tilde{S}} \sum_{1\leq i < l \leq n} \left\lbrace \widetilde{\dot{U}(X_i(t),X_l(t))} \widetilde{V(Y_i(t),Y_l(t))} - \mathbb{E}\left[ \widetilde{\dot{U}(X_i(t),X_l(t))} \widetilde{V(Y_i(t),Y_l(t))} \right] \right\rbrace \stackbin{d}{\longrightarrow} N(0,1).
\end{equation*}\\

\subsection{Simulation details for considered competitors}\label{append:comp_simu}

In this section we remind the scenario A structure of the FLCM algorithm implemented in \cite{GHOSAL2022Score} as well as the form of scenario (B) for the ANFCM testing performance in \cite{Kim2018Additive}. Moreover, we explain how these algorithms are implemented to replicate their results.\\

\subsubsection{Implementation details for FLCM algorithm} 

Under the linear assumption, data $i=1,\dots,n$ is generated as 
\begin{equation*}
	Y_i(t)=\beta_0(t)+X_i(t)\beta_1(t)+\varepsilon_i(t)
\end{equation*}
where $\beta_0(t)=1+2t+t^2$ and $\beta_1(t)=d\cdot t/8$, for $d\geq 0$. The original covariate samples $X_i(\cdot)$ are i.i.d. copies of $X(\cdot)$, where $X(t)=a+b \sqrt{2}\sin(\pi t)+ c  \sqrt{2}\cos(\pi t)$, with $a\sim N(0,1)$, $b\sim N(0,0.85^2)$ and $c\sim N(0,0.70^2)$ independent. It is assumed that the covariate $X_i(t)$ is observed with error, i.e. $W_i(t)=X_i(t)+\delta_{it}$ is getting instead, where $\delta_{it}\sim N(0,0.6^2)$ and changes with every $i$ and $t$. The error process is considered as
\begin{equation*}
	\varepsilon_i(t) = \xi_{i1} \sqrt{2} \cos(\pi t) + \xi_{i2} \sqrt{2} \sin(\pi t) + \xi_{i3t}
\end{equation*}
where $\xi_{i1}\stackbin{iid}{\sim} N(0,2)$, $\xi_{i2}\stackbin{iid}{\sim} N(0,0.75^2)$ and $\xi_{i3t}\stackbin{iid}{\sim} N_\mathcal{T}(0,0.9^2 I_\mathcal{T})$, with $\xi_{i3t}$ being generated as a multivariate normal of dimension $\mathcal{T}$ and these values change with $i$ and $t$.\\

We consider the dense design, taking a total of $\mathcal{T}=81$ equidistant time points in $[0,1]$, being $t_1=0$ and $t_{81}=1$. A Monte Carlo study is carried out using $M=1000$ replicates to measure calibration and power, and p-values are calculated by means of $B=100000$ samples generated under the null hypothesis of no effect ($H_0: \beta_1(t)=0$ for all $t$). Following authors guidelines, the number of basis components considered is $K=7$. In order to measure calibration and power we consider $d=0$ and $d=3,7$, respectively. Besides, we take $n=60,100$ to compare their results with the MDD-based test ones. To implement this algorithm we have used the public code which can be found in \textcolor{blue}{\url{10.1016/j.ecosta.2021. 05.003}}. In particular, we generate the data and use the \texttt{FLCM.test1} function of the \texttt{test.R} script to implement the test.\\

\subsubsection{Implementation details for ANFCM algorithm} 

In the case of the ANFCM algorithm we perform Algorithm 1 of \cite{Kim2018Additive} in hypothesis testing, which translates in testing the nullity of the second additive effect by
\begin{equation*}
	H_0: \mathbb{E} \left[ Y(t) |_{X_1(t)=x_1} \right]=F_0(t)
\end{equation*}

In this scenario, samples are generated verifying the additive assumption by means of
\begin{equation*}
	Y_i(t)= F_0(t)+F_1(X_{1i}(t),t) \quad \text{for}\quad i=1,\dots,n
\end{equation*}
with $F_0(t)=2t+t^2$, $F_1(X_{1i}(t),t)=d\{2\cos(X_1(t)t)\}$ for $d\geq 0$. True covariate is given by $X_1(t)=a_{0}+a_{1}\sqrt{2}\sin(\pi t)+a_{2}\sqrt{2}\cos(\pi t)$ where $a_{0}\sim N\left( 0, \{2^{-0.5}\}^2 \right)$, $a_{j1}\sim N\left( 0, \{0.85\times 2^{-0.5}\}^2 \right)$ and $a_{2}\sim N\left( 0, \{0.7\times 2^{-0.5}\}^2 \right)$. However, it is assumed that covariate is observed with error, in particular we get $W_{1i}=X_{1i}(t)+\delta_{it}$ with $\delta_{it}\sim N(0,0.6^2)$ varying with $i$ and $t$. The considered error process is 
\begin{equation*}
	\varepsilon_i(t) = \xi_{i1} \sqrt{2} \cos(\pi t) + \xi_{i2} \sqrt{2} \sin(\pi t) + \xi_{i3t}
\end{equation*}
where $\xi_{i1}\stackbin{iid}{\sim} N(0,2)$, $\xi_{i2}\stackbin{iid}{\sim} N(0,0.75^2)$ and $\xi_{i3t}\stackbin{iid}{\sim} N_\mathcal{T}(0,0.9^2 I_\mathcal{T})$, with $\xi_{i3t}$ being generated as a multivariate normal of dimension $\mathcal{T}$ and these values change with $i$ and $t$.\\

The dense design scenario is considered with $\mathcal{T}=81$ equidistant time points in $[0,1]$, being $t_1=0$ and $t_{81}=1$. To study its calibration and power behavior a Monte Carlo study is carried out. We employ $M=1000$ replicates to study both, the calibration and power. In this case, p-values are calculated by means of $B=200$ bootstrap samples in all cases. Besides, following \cite{Kim2018Additive} parameters selection, the number of basis components taken is $K=7$. In order to measure calibration and power we test with $d=0$ and $d=3,7$, to simulate under the null and alternative hypothesis, respectively. Besides, we take $n=60,100$ to compare their results with the MDD-based test ones. We have found the code available in \textcolor{blue}{\url{https://www4.stat.ncsu.edu/~maity/software.html}} and we borrowed it to reproduce the ANFCM simulations. Specifically, we make use of the \texttt{anova.datagen} function of the \texttt{datagenALL.R}\footnote{We have adapted the code to correctly generate $X1(t)$ and $Y(t)$. In particular, in the \texttt{X} function, we have changed \texttt{X.list[[q]]  =2\textasciicircum (1-q)*(a0\%*\%t(ones)+a1\%*\%t(phi1)+a2\%*\%t(phi2))} for the expression \texttt{X.list[[q]] = (a0\%*\%t(ones)+a1\%*\%t(phi1)+a2\%*\%t(phi2))/sqrt(2\textasciicircum(q-1))} to correct a typo. Besides, it is needed to change \texttt{F.anova2 = function(x1,x2,t,d){2*t+t\textasciicircum2+x1*sin(pi*t)/4+d*2*cos(x2*t)}} by \texttt{F.anova2 = function(x2,t,d){2*t+t\textasciicircum2+d*2*cos(x2*t)}} as well as \texttt{Fanova = F(Xeval[[1]],Xeval[[2]],trep,d)} by \texttt{Fanova = F(Xeval[[2]],trep,d)} in function \texttt{anova.datagen} to correctly define the modified version. } script to generate the data and apply \texttt{test.anova} function of the \texttt{test.R} script to implement the algorithm, using now \texttt{list(null.data.dn\$Weval[[2]])}.

\subsection{Graphics}\label{append:extra_simu}

\begin{figure}[htb]\centering
	\includegraphics[width=\linewidth]{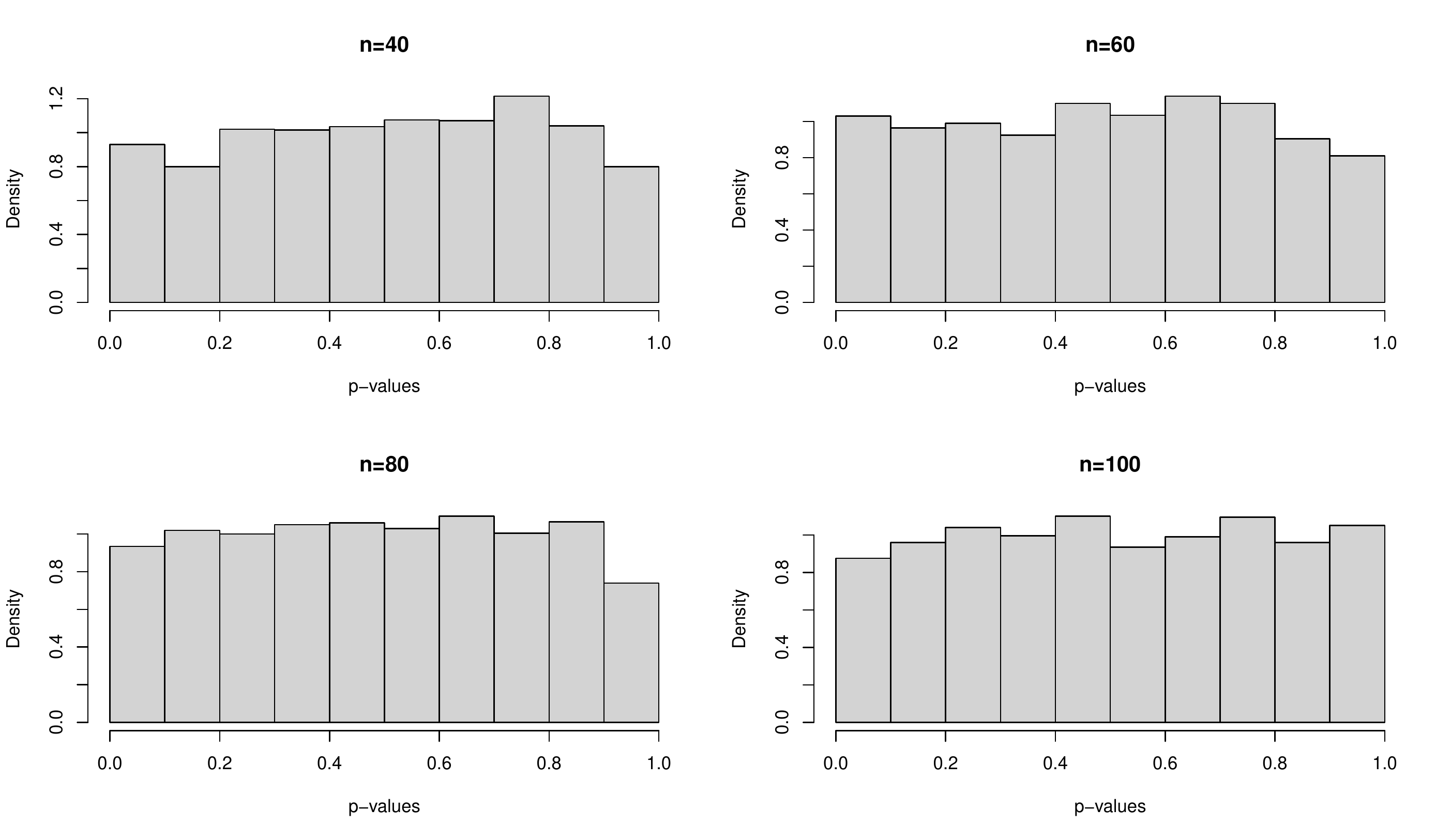}
	\caption{\label{p_values_dist_MDD_CM_global_boot_U} Histograms of the p-values of the test statistics under $H_0$ using the wild bootstrap critical value for some values of $n$ in Scenario A.}
\end{figure}

\begin{figure}[htb]\centering
	\includegraphics[width=\linewidth]{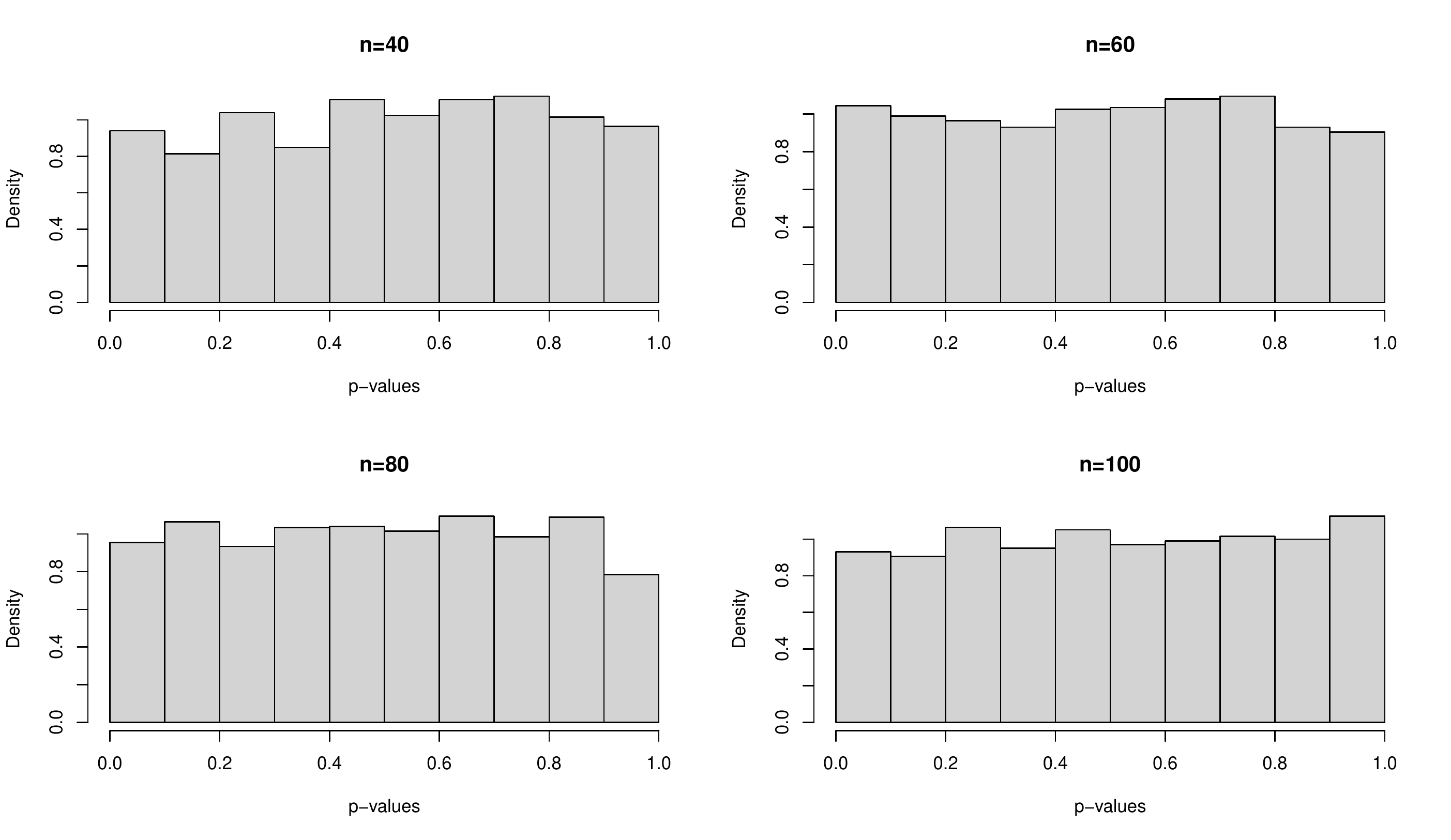}
	\caption{\label{p_values_dist_MDD_CM_global_boot_NL} Histograms of the p-values of the test statistics under $H_0$ using the wild bootstrap critical value for some values of $n$ in Scenario B.}
\end{figure}

\begin{figure}[htb]\centering
	\includegraphics[width=\linewidth]{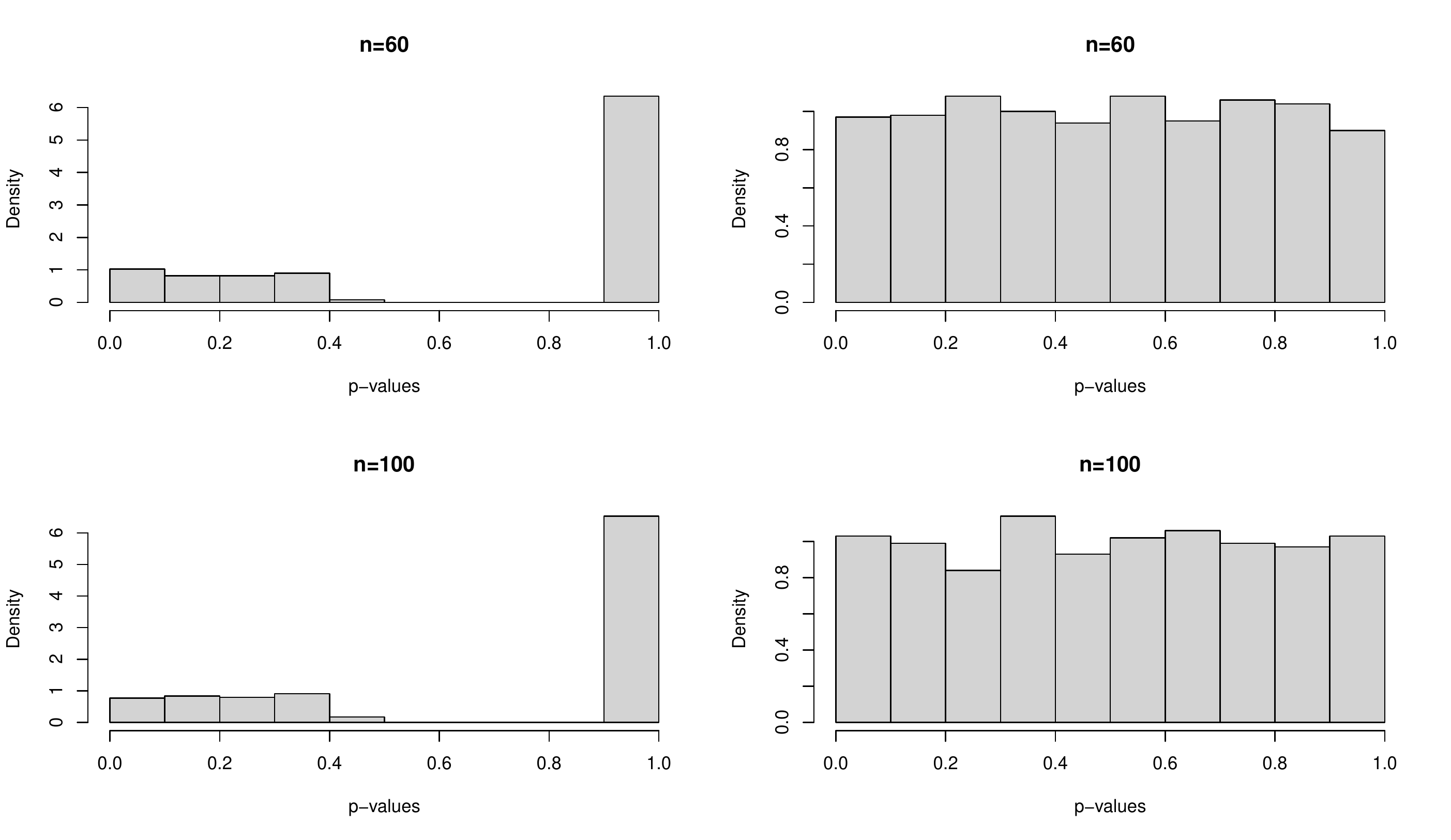}
	\caption{\label{p_values_comparison_FLCM_MDD.pdf} Histograms of the test statistics p-values under $H_0$ for the FLCM (left column) and MDD (right column) methods in scenario A of \cite{GHOSAL2022Score}.}
\end{figure}

\begin{figure}[htb]\centering
	\includegraphics[width=\linewidth]{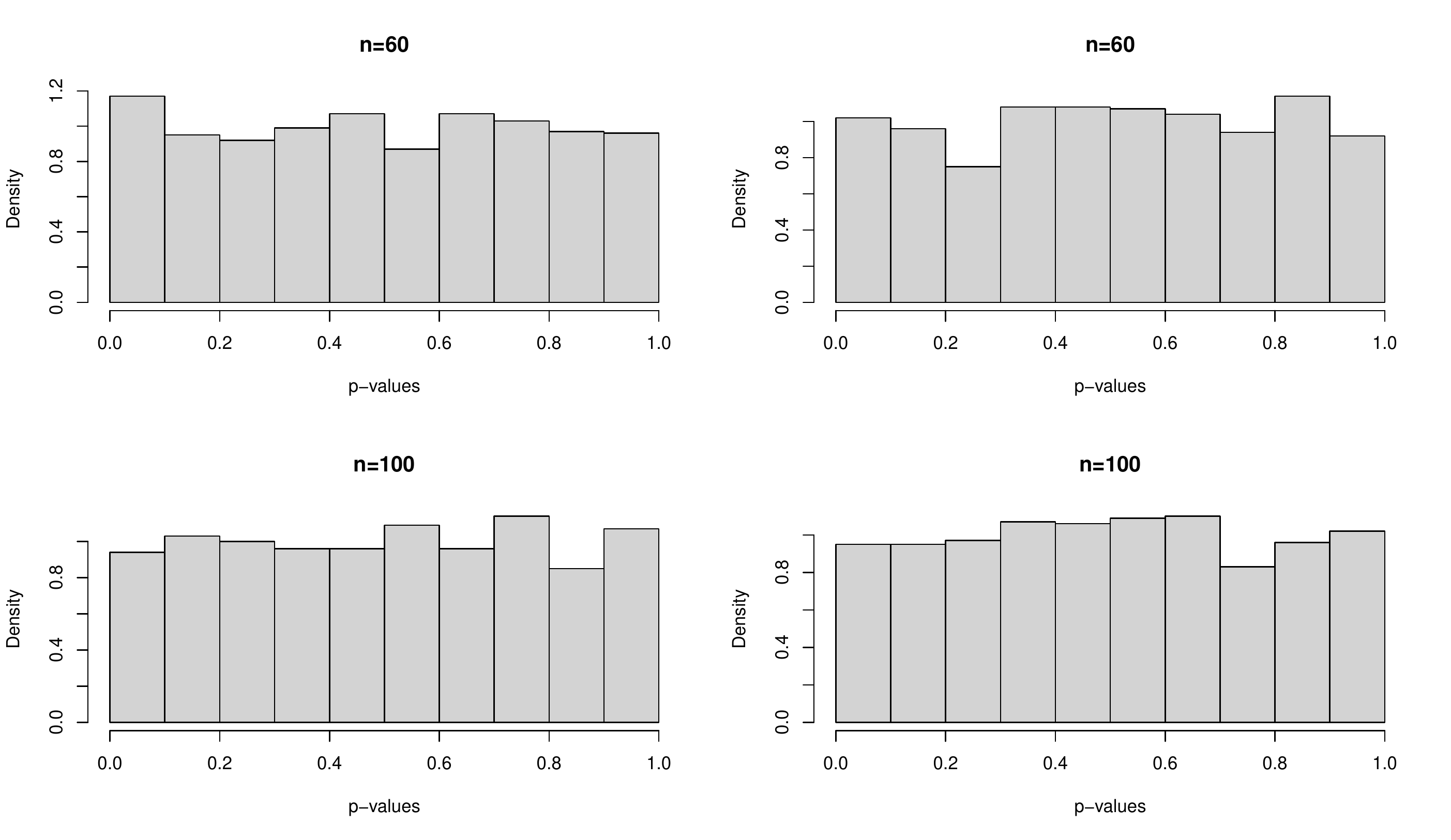}
	\caption{\label{p_values_comparison_ANFCM_MDD.pdf} Histograms of the test statistics p-values under $H_0$ for the ANFCM (left column) and MDD (right column) methods in modified scenario B of \cite{Kim2018Additive}.}
\end{figure}

\end{document}